\documentclass{article}

\usepackage{PRIMEarxiv}

\usepackage[utf8]{inputenc} 
\usepackage[T1]{fontenc}    
\usepackage{hyperref}       
\usepackage{url}            
\usepackage{booktabs}       
\usepackage{amsfonts}       
\usepackage{nicefrac}       
\usepackage{microtype}      
\usepackage{lipsum}
\usepackage{fancyhdr}       
\usepackage{graphicx}       
\graphicspath{{media/}}     
\usepackage{acronym}
\usepackage{subfigure}

\title{Smart Radio Environments}

\author{
  Gabriele Gradoni \\
  School of Mathematical Sciences\\ 
  Department of Electrical and Electronic Engineering\\
  University of Nottingham \\
  Nottingham\\
  United Kingdom\\
  \texttt{gabriele.gradoni@nottingham.ac.uk} \\
   \And
  Marco Di Renzo \\
  Laboratoire des Signaux et Syst\'emes\\ 
  CNRS, CentraleSup\'elec\\ 
  Université Paris-Saclay\\ 
  Gif-sur-Yvette\\
  Paris\\ 
  France\\
  \texttt{marco.direnzo@l2s.centralesupelec.fr} \\
}

\begin{document}
\maketitle

\begin{abstract}
This Roadmap takes the reader on a journey through the research in electromagnetic wave propagation control via reconfigurable intelligent surfaces. Meta-surface modelling and design methods are reviewed along with physical realisation techniques. Several wireless applications are discussed, including beam-forming, focusing, imaging, localisation, and sensing, some rooted in novel architectures for future mobile communications networks towards 6G.
\end{abstract}


\section{Introduction}
Future mobile wireless network technologies seek to extend radio signal coverage and boost communication data-rates of an ever increasing number of mobile subscribers.  
It is estimated that the number of mobile terminals worldwide will almost double the global population by 2025. 
The recent development of reconfigurable intelligent surfaces (RISs) provides a way forward to support the ambition of mobile network technologies in the transition from 5G to 6G. 
Inherently, it is believed that properly designed RISs are amenable to integration within existing 4G/5G network infrastructures. 
Several international research projects are currently underway to substantiate this belief, thus creating a multidisciplinary arena where physicists, mathematicians, engineers, and computer scientists, can collaborate to address the most pressing challenges generated by new RIS-based network architectures. 
While metamaterials (MTS) constitute the physical backbone of RISs, their optimization and orchestration as a network resource/service call for advanced electromagnetic (EM) modelling techniques, intertwined with modern artificial intelligence (AI) and machine learning (ML) methods. 
This roadmap reviews the basic concepts, discusses open problems, and investigates future solutions, from the perspective of prominent scientists actively researcheing on RISs.

\subsection{Roadmap in detail}
The roadmap is organised in three sections that mirror the parallel effort of the electromagnetic and physics communities to develop MTS that have been recently included in advanced solutions for beyond 5G/6G reconfigurable mobile networks.   
\begin{itemize}
\item \textbf{EM Modelling and Design:}
\emph{Diaz-Rubio} and \emph{Tretyakov} identify gaps between EM theory and design of MTS.
\emph{Caloz} discusses modelling assisted design for near-field wave and quantum manipulation, also touching on quantum computing. 
\emph{Gradoni} and \emph{Peng} show how to find the optimal state of very large RISs via Ising models and quantum annealing.
\emph{Alu} introduces metagratings for reflections at extreme angles and highly efficient focusing. 
\item \textbf{Physical Realization and EM Environment:}
\emph{Lerosey} and \emph{Fink} look back at the pioneering achievement of a RIS prototype and unfold the road for the realization of self-adaptive RISs. 
\emph{Galdi} and \emph{Cui} present challenges in the modelling of space-time coding digital MTS for an all-EM modulation scheme. 
\emph{Frazier} and \emph{Anlage} discuss the integration of RISs within complex cavities and the need of a ML-assisted wavefront shaping.
\emph{Salucci} and \emph{Massa} extend wave control to large scale EM environments showing that arbitrary sources can be designed by inverse methods including smart objects. 
\item \textbf{Wireless System Applications:} 
\emph{Cheng} and \emph{Cui} survey the use of RISs in wireless communications and explain the importance of low complexity coding MTS in 6G wireless newtorks. 
\emph{Wang} and \emph{Jin} tackle the fundamental challenges in signal detection, channel estimation, and beamforming for RIS assisted wireless networks via AI methods.
\emph{Dardari} and \emph{Decarli} elucidate the need of better EM and channel modelling of RIS assisted systems to support the realization of holographic radios. 
\emph{Yurdusevem} and \emph{Matthaiou} offer a perspective on the use of RIS apertures to exploit holography in direction of arrival estimation and beam synthesis. 
\emph{Kenney} and \emph{Gordon} envisage compact RIS assisted imaging solutions in the optical domain as an innovative multimode IoT device. 
\emph{Georgiou} and \emph{Nguyen} examine the role of RISs in the enhanchement of indoor and underground localization technologies for 6G. 
\emph{Martini} and \emph{Maci} propose an all-EM wireless sensing solution based on the manipulation of surface waves propagating through MTS to reduce network complexity. 
\emph{Wakatsuchi} and \emph{Phang} elaborate on the use of RISs to sense waveforms and their use in wireless communications. 
\end{itemize}

\newpage

\section*{\fontsize{20}{25}\selectfont Scattering from reconfigurable metasurface panels} 
\subsection*{Ana Díaz-Rubio and Sergei A. Tretyakov*} 
\subsubsection*{\textsuperscript{1}\textit{Department of  of Electronics and Nanoengineering, Aalto University, Espoo, Finland}} 
\subsubsection*{*\textbf{Corresponding author}: sergei.tretyakov ``at'' aalto.fi} 
\vspace{5mm}

\paragraph{Emerging challenges}
Recently, many researchers have been studying potential improvements of wireless telecommunication systems with the use of reconfigurable intelligent metasurfaces. 
The governing idea is that making some parts of the propagation environment tunable and adjustable, one can optimize it together with the optimization of  the transmitters and receivers. 
When experts in communication theory, signal processing, and machine learning consider wireless propagation channels in presence of reconfigurable metasurfaces, they need to model reflected and scattered fields produced by anomalously reflecting metasurfaces. The wast majority of such studies are based on the assumption that every ``point'' at the metasurface plane or every ``unit cell'' of the metasurface is characterized by a certain reflection coefficient 
\begin{equation}
    \Gamma(x,y)=|\Gamma(x,y)|e^{j\Phi(x,y)}
    \label{local_Gamma}
\end{equation}
that can be tuned at will for any coordinate $x,y$ at the metasurface (see, e.g., \cite{tut} and references therein).

\begin{figure}[h]\centering
	\includegraphics[width=0.95\linewidth]{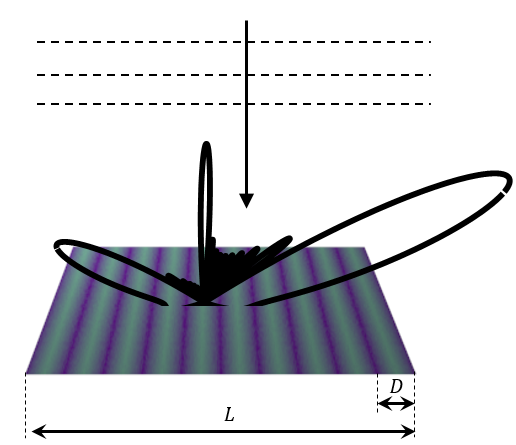}
	\caption{Schematic representation of the scattering produced by an anomalous reflector designed with a linear phase gradient of local reflection coefficient. The use of model \ref{local_Gamma} predicts only one reflected beam with a large error in the estimated amplitude of reflection.}
	\label{fig1T}
\end{figure}


Propagation over the air from the transmitter to every point of the metasurface and from these points to the receiver are similarly modeled by the corresponding complex-valued factors, and, finally, the function $\Gamma(x,y)$ is optimized with the goal to increase the signal strength at the receiver position.

However, the model described by Eq.~\ref{local_Gamma} is valid only in the assumption that the metasurface is approximately uniform at the wavelength scale (in the theory of diffraction this model is called \emph{physical optics}, e.g.,  \cite{Osipov}). For anomalously reflecting metasurfaces, this assumption holds only for small tilts of the reflected wave with respect to the specular-reflection angle. For modeling of reflections from ``unit cells'', this model is not useful, because  single array elements scatter in all directions, and such reflection coefficient is not meaningful. Moreover,  unit-cell response models completely neglect field coupling between cells, while the design of effective anomalous reflectors is based on proper engineering of this coupling (e.g., \cite{1,2,3}). On the other hand, the electromagnetic theory of anomalous reflectors (especially finite-size panels) is not yet properly developed, and the results often come in a form not directly suitable for inclusion in channel models or ray-tracing algorithms. 

\paragraph{Future developments to satisfy these challenges} 
Attending to the current demands on the development of smart radio environments, the first challenge is to find appropriate models for metasurface response to external illuminations and  calculating reflected and scattered fields (including both reflection and transmission) of  finite-sized periodic metasurface. Most of known approaches to calculation of reflection from finite-sized metasurfaces are based on the local reflection coefficient model \ref{local_Gamma} and, as it was discussed in the introduction, do  not properly cover the complexity of the problem. 
A possible approach to address this challenge  merges some ideas of the physical optics (but using nonlocal expressions for the induced currents) and the theory of diffraction by gratings (because periodical structures create multiple diffracted beams) \cite{our}. 
Figure~\ref{fig1T} schematically represents the scattering pattern of an anomalous reflector designed with a linear phase gradient of the local reflection coefficient when we consider the energy coupling between the incident wave and  different propagating diffracted modes.
Although this model properly accounts for scattering into all propagating modes, the model assumes homogeneous distribution of the equivalent currents over the metasurface neglecting the effect of the discontinuities at the borders. This simplification has an impact on the accuracy of the method for calculating scattering from small-sized metasurfaces. 

\begin{figure}[h]\centering
	\includegraphics[width=0.95\linewidth]{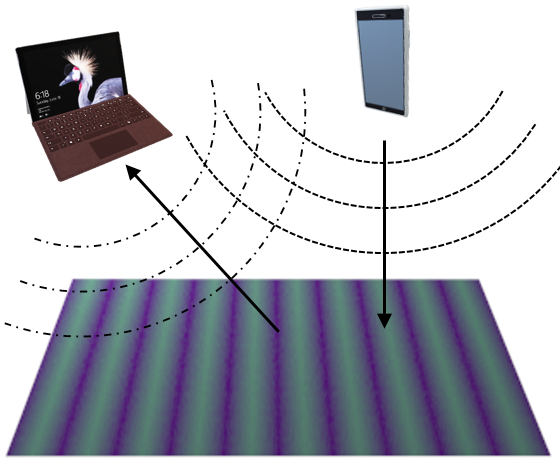}
	\caption{Illumination of an anomalous reflective metasurface by a receiver located in close proximity to the metasurface. }
	\label{fig2T}
\end{figure}

Generally, metasurfaces have been studied assuming homogeneous plane-wave illumination from a certain direction. This assumption is valid if the source is far from the metasurface. In real-world application and especially in in-doors applications, this condition may not always be valid. 
As it is shown in Fig.~\ref{fig2T}, even though metasurfaces are typically periodic systems, the proximity of the transmitter to the metasurface cause a spatial variation of the incident field that wrecks the periodicity of system. 
For this reason, appropriate models should be applicable to scenarios where the metasurface is aperiodic and inhomogeneously illuminated.  
Developing of such models is challenging because  the scattering produced by the metasurface does not follow the theory of diffraction by gratings.

It is expected that future communication systems will benefit from the full potential of inhomogeneous and time-varying metasurfaces with engineered and adaptive spatial and temporal properties. It is obvious that reconfigurable metasurfaces whose operation modes can be changed according to the demands of the environment have an interesting role to play to control the propagation channel. However, one can envision even more exotic functionalities, from non-reciprocity to loss compensation, by exploiting the possibilities offered by time-modulated metasurfaces.

\paragraph{Conclusion}
We see that there is a deep gap between the electromagnetic theory and design of reconfigurable metasurfaces that leads to appropriate physical models of their response and the communication-theory models of propagation channels in presence of such new devices. Without bridging this gap, many results of advanced studies of channel optimization will produce estimations of performance of structures that are not realizable in principle. There are several publications that try to fill this gap, e.g. \cite{Marco,Gradoni,our}, but there is still a long road ahead.

It is also important to note that experimental work on developing reconfigurable metasurfaces for telecommunications lags behind theoretical contributions. There has been only a few conceptual demonstrations so far, mainly for discrete (usually only two states) tuning of unit cells, e.g. \cite{Cui}. Design of a fully  tunable (with some limitations, naturally) metasurface can be found in \cite{design}, but experimental results are not yet available. 

Finally, we note that the current efforts are focused on reconfigurable structures whose functionality is fixed between the acts of reconfiguration. However, continuous coherent tuning can offer totally new possibilities for control of wave reflection and transmission, e.g. \cite{Xuchen}. We expect that these possibilities will be exploited in future wireless communication systems.

\phantomsection
\section*{Acknowledgment}
This work was supported in part by the European Commission through the H2020 ARIADNE project under grant 871464 and the Academy of Finland under grants 330260 and 330957.


\newpage

\section*{\fontsize{20}{25}\selectfont Physics and Modeling of Metasurfaces} 
\subsection*{Christophe Caloz\textsuperscript{1}*} 
\subsubsection*{\textsuperscript{1}\textit{
		META Research Group, KU Leuven, Belgium}} 

\subsubsection*{*\textbf{Corresponding author}: christophe.caloz@kuleuven.be} 
\vspace{5mm}

\begin{figure}[ht]\centering
\includegraphics[width = .95\linewidth]{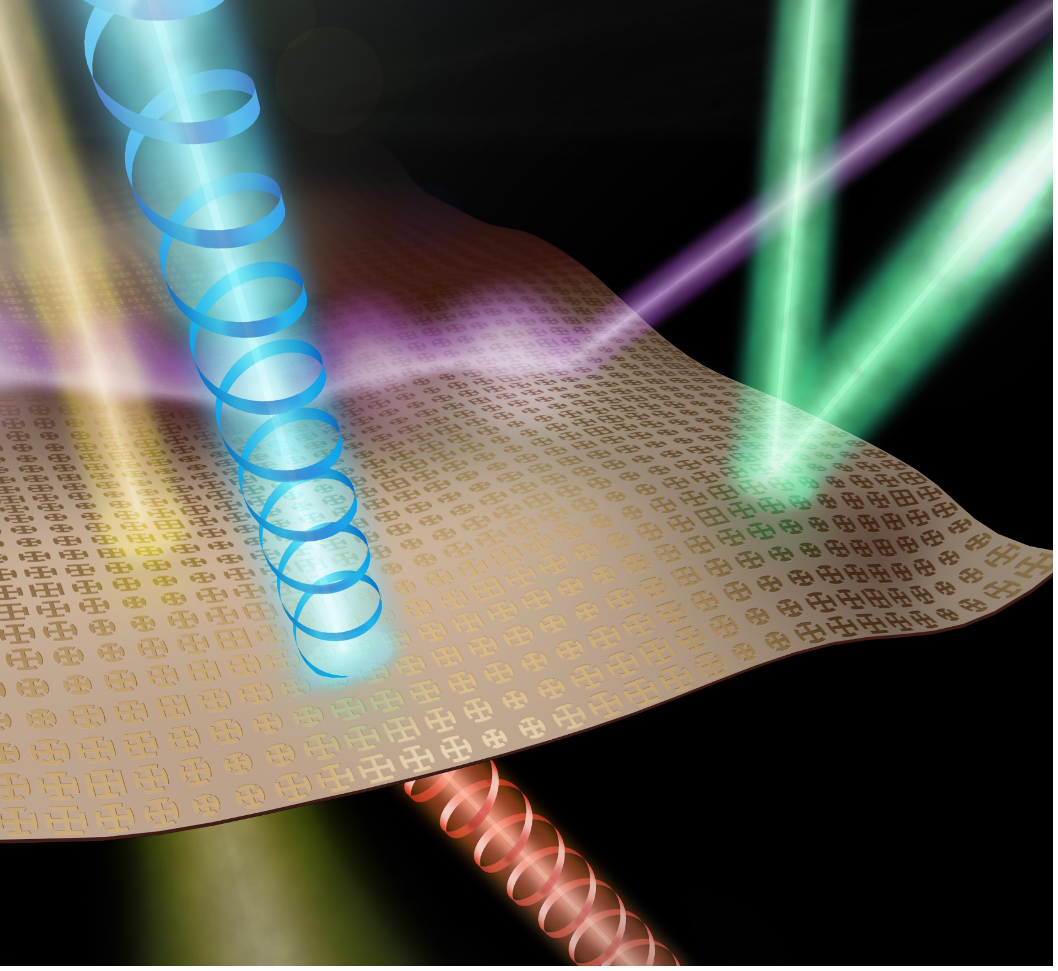}
\caption{\label{fig:overview} \bf Artistic representation of a generic metasurface, performing a diversity of sophisticated electromagnetic wave transformations.}
\end{figure}

\paragraph{Introduction}
Metasurfaces (Fig. \ref{fig:overview}) are electrically thin sheets of subwavelength artificial particles - or 
metaparticles - arranged in a periodic lattice and designed to transform 
electromagnetic waves\footnote{Although we focus here on electromagnetic metasurfaces, metasurfaces may be of different types, depending on the type of waves that they manipulate; they can also be acoustic, mechanical, and gravitational.} according to specifications \cite{Ref1}. They may be 
considered as the 2D counterparts of voluminal metamaterials, with the 
advantages of lower form factor, smaller loss and easier fabrication, or 
as generalizations of frequency or polarization selective surfaces 
spatial light modulators, with the advantage of drastically diversified 
properties and enhanced capabilities across the entire spectrum from 
microwave frequencies to optical wavelengths via the terahertz regime.
Metasurfaces have emerged as a novel 
paradigm in science and technology over the past decade. Given their 
simple configuration, one may wonder why they did not flourish much 
earlier. The reason is that their macroscopic simplicity conceals a 
formidable complexity and diversity at the microscopic level of their 
metaparticles, especially when the structure is multilayered, as is 
typically the case for surface impedance matching, and nonuniform, i.e., 
involving particles of different sizes and/or shapes. Mastering such a 
jungle of parameters necessitated a deep understanding of artificial 
materials, which required itself about one century and a half of 
intensive research on complex media, artificial dielectrics and modern 
metamaterials. The advance that decisively promoted metasurfaces to the 
rank of a new paradigm has been the progressive understanding on how to 
implement bianisotropy on the two-dimensional platform.

\paragraph{Current and Future Challenges}
\textbf{Bianisotropy.} 
Bianisotropy is characterized by 
the constitutive relations $D =\varepsilon _{0}({I}+{\chi 
}_{ee})\cdot E+\frac{1}{c}{\chi }_{em}\cdot H$ and $B={\chi 
}_{me}\cdot E+\mu _{0}({I}+{\chi }_{mm})\cdot H$ [1]. 
It is a type electromagnetic response that involves 36 complex medium 
parameters (4 tensors of dimension 33), some of which are typically 
interrelated by fundamental medium properties, such as reciprocity, 
loss-gain, chirality, gyrotropy and birefringence. Historically, 
bianisotropy first appeared, in partial tensorial form, in media moving 
with respect to a rest frame \cite{Ref2}, and it was generalized to its full 
(36-parameter) tensorial response in the 70's\footnote{Bianisotropy is not the most general form of medium response (which would include increasing-order spatial vectorial derivatives of $E$ and $H$), but only a particular case of weak spatial dispersion \cite{Ref4}. However, it generally models metamaterials adequately because their subwavelength metaparticle feature, leading to material homogeneity, typically prevents the onset of higher-order spatial dispersive effects.} \cite{Ref3}. However, 
general bianisotropic materials are not widely available in nature, and 
their implementation in artificial form has remained elusive, due to the 
prohibitive challenge posed by their fabrication complexity. 
Metasurfaces, thanks to their much simpler, two-dimensional 
configuration, have allowed, for the first time, practical access to the 
fantastic diversity of full bianisotropy, and subsequently led to a 
myriad of novel electromagnetic effects and applications \cite{Ref1}.\\
\textbf{GSTC.}  
Given their deeply subwavelength 
thickness, metasurfaces do generally\footnote{This constraint has recently been relaxed in several metasurface designs, but the GSTCs can then still be sometimes used in terms of effective sheet conditions.} not support normal Fabry-Perot 
resonances, and can therefore be perfectly modeled by Generalized Sheet 
Transition Conditions (GSTCs). GSTCs are generalizations of textbook 
boundary conditions, which include the addition of \textit{surface }
polarization currents that model the response of the metaparticles and 
their intercoupling in the overall metasurface structure \cite{Ref1,Ref5,Ref6,Ref7}. 
They have the general form \cite{Ref1}:
$$
{n}\times \Delta H =\frac{\partial 
}{\partial t}(\varepsilon _{0}{\chi }_{ee}E_{av}+\frac{1}{c}{\chi 
}_{em}H_{av})+{n}\times \nabla M_{z},
$$
$$
\Delta E \times {n}=\mu 
_{0}\frac{\partial }{\partial t}({\chi }_{mm}H_{av}+\frac{1}{\eta }{\chi 
}_{me}E_{av})+\frac{1}{\varepsilon _{0}}{n}\times \nabla MP_{z},
$$
In these relations, the $\Delta $ and $av$ terms represent the differences and averages, respectively, of 
the electric and magnetic fields at either side of the metasurface, with ${n}$ being the unit vector normal to it.\\ 
\textbf{Design.} 
In a synthesis problem, where the task is 
to design a metasurface that performs a given electromagnetic wave 
transformation, the field quantities are known, and one can then solve 
the above equations for the susceptibility tensors. In linear 
time-invariant metasurfaces, the convolution products in these relations 
reduce to multiplication products, and these tensors can often be found 
in closed form \cite{Ref1}. New full-wave algorithms have been developed to 
analyze metasurfaces characterized by the bianisotropic tensor 
susceptibilities obtained from this procedure \cite{Ref8}. Once the 
metasurface described by these continuous tensorial susceptibility 
functions has been synthesized and analyzed, it is discretized in 
subwavelength periodic cells in each of which a specific metaparticle 
(size and shape) is found via a parametric map between the theoretical 
scattering response associated with the susceptibility model and the 
full-wave simulated scattering response of specific metaparticle 
structures \cite{Ref1}.\\
\textbf{Applications.} 
The unique and systematic 
capabilities of metasurfaces to control the phase, the magnitude, and 
the polarization of waves in terms of 36 bianisotropic parameters, 
combined with the recent availability of powerful design techniques 
based on GSTCs, has already led to uncountable applications at the time 
of this writing \cite{Ref1}. They are too numerous to be exhaustively 
mentioned here but they include generalized reflection and reflection, 
all kinds of other wavefront manipulations, generalized Brewster and 
Fresnel refraction, agile polarization processing for gyrotropy, 
birefringence and polarimetry, sophisticated spatial and magnetless 
nonreciprocity, multi-wave transformation, simultaneous spatial and 
temporal spectral processing, spatial nonlinearity boosting, imaging and 
holography, spin and orbital angular momentum transformation, 
angle-independent absorption, wireless communication smart-panel 
scattering and processing, camouflaging and cloaking, thermal radiation 
management, nanoparticle and solar sail optical force tailoring, power 
harvesting, quantum system engineering, analog computing; references to 
most of these are found in the review papers \cite{Ref9}-\cite{Ref12}.

\paragraph{Advances in Science and Technology to Meet Challenges}
At this time, the 
field of metasurfaces faces several theoretical and technological 
challenges, as well as related opportunities. I shall try to identify 
here some of these challenges and opportunities: \underline{Theory}: 
1) The brute-force parametric-mapping technique mentioned above for the 
synthesis of specific metaparticles from synthesized tensorial 
susceptibility functions is slow and tedious; the development of smarter 
approaches, possibly drawing from topology and group theories and 
machine-learning artificial intelligence, would be highly welcome to 
provide more efficient design strategies. 2) As mentioned above, 
bianisotropy represents a low spatial dispersion model; exploiting 
higher-order spatial dispersion could both lead to more accurate 
modeling of current metasurfaces and enable metasurfaces with greater 
diversity transformations in mesoscopic-regime structures. 
3) Metasurface near-field manipulations have been very little considered 
to date; their investigation could enable a novel range of applications, 
for instance in microscopy and Casimir-force engineering. 4) Despite the 
current blossoming of quantum technologies, few efforts have been 
dedicated to the control of quantum systems by metasurfaces, not to 
mention the development actual quantum metasurfaces; research efforts in 
this area will certainly bring novel effects and devices for quantum 
sensing, computing and metrology. \underline{Technology}: 1) While 
many metasurfaces can be easily fabricated in the microwave regime, 
their implementation at optical wavelengths is still often challenging, 
particularly in configurations involving multiple layers and fine 
features; much effort is currently being deployed in this area, 
particularly using all-dielectric and CMOS platform approaches. 
2) Nonreciprocity has been recently realized to be particularly 
desirable to come in magnetless format for compatibility with integrated 
circuit technology and spatial processing, particularly in 
transistor-loaded metaparticle technology \cite{Ref13}; however, serious 
efforts are still required to develop efficient nonreciprocal 
metaparticles in the microwave regime, and to find a substitute to 
transistors - or develop an optical transistor \cite{Ref14} - in the optical 
regime\footnote{Time-modulated nonreciprocity does currently offer a better option than transistor nonreciprocity technology because of the speed limitation of the electronics required for modulation (while record transistors have recently passed the symbolic operation limit of 1 THz).}. 
3) The addition of time modulation to space modulation towards 
spacetime-modulated \textit{metamaterials} represents particularly 
vast potential of scientific and technological innovation \cite{Ref15,Ref16}; 
although this potential has been recently discussed in metasurface 
structures \cite{Ref17}, the implementation of this concept is still 
enormous challenging, and will require novel technological approaches 
for the "\,RF modulation" of the metasurfaces. 4) Finally, smart 
metasurfaces, serving for instance in wireless communication systems for 
"\,fog-type" local processing to reduce the cluttering of servers on the 
internet of things, will require fully programmable tunable and active 
metasurfaces acting as local distributed processors.


\newpage

\section*{\fontsize{19}{23}\selectfont On the Role of Quantum Optimization in Reconfigurable Wireless Environments} 
\subsection*{Gabriele Gradoni\textsuperscript{1} and Zhen Peng\textsuperscript{2}*} 
\subsubsection*{\textsuperscript{1}\textit{School of Mathematical Sciences and Department of Electrical and Electronic Engineering,  
University of Nottingham, University Park NG72RD, United Kingdom}} 
\subsubsection*{\textsuperscript{2}\textit{Department of Electrical and Computer Engineering, University of Illinois at Urbana-Champaign,  Illinois, USA}} 
\subsubsection*{*\textbf{Corresponding author}: zvpeng@illinois.edu} 
\vspace{5mm}


\paragraph{Motivation:}
The reconfigurable intelligent surface (RIS) is emerging as a key technology for the next-generation of mobile communication network. The general goal is to turn the wireless environment into a smart/reconfigurable space that plays an active role in the wireless communication performance \cite{RISE-6G-2021}.
Going beyond 5G and entering 6G, it is envisaged that large-scale, distributed RIS devices may be deployed at the surface of interacting  objects, e.,g. wall, windows, furniture, in the propagation channel. 
A joint optimization of wireless endpoints and distributed RISs would lead to a dynamically programmable and customized wireless environment, with a goal of providing enhanced coverage with high energy efficiency and supporting ultra-fast and seamless connectivity. 
To harness the full potential of RIS-enabled smart radio environment, we need to rapidly optimize the states of RIS devices with prescribed objective functions. 
This constitutes a substantial computational task both in the physical and network layer of wireless communication, due to the enormous number of available degrees of freedom (DOFs).

 

\paragraph{Introduction to Quantum Optimization:}
Whereas some recent researches pursue the direction of artificial intelligence and deep learning, we elaborate the role of quantum computing (QC) to provide a scalable approach that overcomes the computational optimization complexity. 
In recent years, the remarkable progress made in QC hardware has defined a new, Noisy Intermediate-Scale Quantum (NISQ), QC era.
By exploiting fundamental properties of quantum mechanics, these QC systems have reach the potential to deliver orders of magnitude in the speedup against classical computing hardware for solving hard problems. 
Here, we focus on quantum combinatorial optimization algorithms, which run on NISQ devices to search for an optimal solution over all the combinatorial states of RIS elements. 
One well-known example is the Grover Adaptive Search algorithm \cite{Grover1997} with provable quantum speedup, which has been recently applied to Constrained Polynomial Binary Optimization (CPBO) problems.
Another representative case is the quantum approximate optimization algorithm (QAOA) \cite{Farhi2014} utilizing unitary operators and quantum superposition. 
It is noted that both above-mentioned algorithms run on universal gate-model quantum computers. There is another 
specialized analog computer, so-called quantum annealer (QA), which belongs to the adiabatic quantum computing (AQC) regime \cite{QA-Book}. 
In what follows, we review the current literature on quantum optimisation of smart/reconfigurable wireless systems and discuss open problems in this area of research, with an outlook into possible solutions to address the most pressing challenges to enable large scale 6G network design.

\paragraph{State of the art:} 
Recently, paradigms based on QC have attracted the interest of network operators as a promising platform to perform flexible optimisation of dense networks that serve an increasing number of users. 
In this context, QC is expected to achieve unprecedented low latency reconfiguration of base stations, as well as more efficient resource scheduling policies. 
A review article on quantum search algorithms in wireless applications can be found in \cite{Botsinis2019}.
In September 2018, a partnership between British Telecommunications and academic partners in the United Kingdom \cite{Popa} has set the sea to use QC in telecommunications. In particular, QA has helped planning the network from system level to optimise network resources, which typically involve hard (NP and P hard) optimisation problems. 
For example, cell user channel allocation and half-duplex network optimisation are known to be NP hard problems. 
In February 2020, Telecom Italia Mobile (TIM) was the first mobile operator to adopt QC approaches in planning of 4.5G and 5G networks \cite{TIM}. 
In particular, the quadratic unconstrained binary optimization (QUBO)  algorithm was implemented on a D-Wave's 2000QTM quantum computer to perform real-time network configuration ten times faster than traditional optimization algorithms. 
This paradigm already provides increased quality of service in Voice over LTE (VoLTE) and is believed to be important in future self-organising networks (SONs). 
More recently, the QA was also used in the vector perturbation precoding for large MIMO systems in downlink \cite{Venturelli2021}. 

\begin{figure}[h]
\begin{center}
\noindent
  \includegraphics[width=\textwidth]{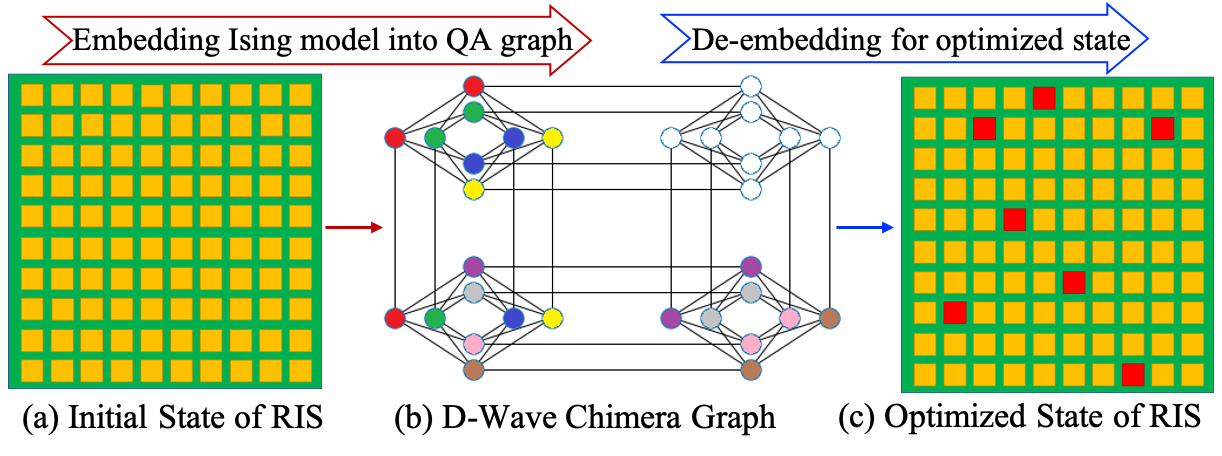}
  \caption{Embedding RIS Ising model into QA Hardware.}\label{Embedding}
  \vspace{-1.0cm}
\end{center}
\end{figure}

Very recently, researchers have drawn their attention to fuse electromagnetic (EM) models with quantum optimization, a research that is gradually defining a new field on quantum optimization of RIS assisted wireless networks. 
In this new field, statistical physics is playing an important role in linking the multi-element, distributed devices, wireless problem to a physical formulation that can be tackled efficiently with novel QC architectures. 
As an example, in statistical mechanics, the Ising model is widely used to describe the spin state of arrays of quantum particles. In \cite{Peng2021}, the authors dwell on this analogy and develop an Ising model for the RISs with prescribed scattering profile. The spin-like degrees of freedom are offered by the discrete phase values in tunable RIS elements. The EM wave energy is then used to calculate the Ising Hamiltonian. 
Subsequently, the Ising model is compiled into a physical QA hardware, the D-Wave 2000Q (DW2Q) QA device, through a process of embedding and de-embedding, shown in Fig. \ref{Embedding}. The configuration time for simultaneous beam- and null-forming of large RISs is achieved within a few milliseconds, which is much less than the channel coherence time in a dynamic radio environment.

  \vspace{-0.24cm}

\paragraph{Emerging challenges:}
In the future, we will face a plethora of challenges pertaining two highly interdisciplinary research areas. 
The first area is the holistic wave modeling and design of joint RIS devices and propagation environments. 
The interaction between RIS and random fields needs an in-depth study that captures the multiple reflections originating within a confined wireless environment. This study requires the formulation as a boundary value problem (BVP) that can be treated with statistical methods developed in the areas of wave and quantum chaos. More specifically, the random RIS scattered field can be obtained from the BVP via semiclassical analysis methods and random matrix theory.

The second area is the global optimization of large networks of RIS devices operating in large scale environments, important for the orchestration of multiple RIS devices. 
More accurate EM models, including multipath fading and near field coupling, will need to be mapped into Ising Hamiltonians for advanced design with multi level phases of the RIS unit cell. 
The marriage of electromagnetism and statistical mechanics for the physical layer optimization represents a promising way forward for reconfigurable environments.
Research in these two areas is of crucial importance in both the planning and the dynamic reconfiguration of mobile networks.  Eventually, the evolution into SONs with quantum computing capabilities will pave the way to real-time network reconfiguration assisted by a Quantum Digital Twin (QDT).

\paragraph{Future developments:}
One research direction is to enhance the fidelity of the quantum-suitable RIS model. To increase the angular resolution of reflected beam, three and four bit codebooks for the reflection phase of RIS unit cells will be formulated in terms of Ising spin variables. 
Multi-level optimization will be considered for very large (or multiple) RIS devices via extension of antenna sub-array design schemes, such that a single large RIS can be applied for multiple access point (AP) multiuser reflective beamforming. Finally, the mutual coupling between RIS elements can be approximated by  impedance matrices of small dipoles with tunable loads and captured in a MIMO channel model that can be used for network design. 
Another research direction will seek to improve the modeling fidelity of the propagation environment. The near-field RIS optimization will be studied to extend the use case scenarios of RISs from far-field to near-field (Fresnel region and near the Fraunhofer distance). Configuring RIS in multi-path channel and under random, e.g., reverberating, fields will also be tackled via stochastic Green function approaches to account for the effect of multi-path fading onto the global optimisation of RIS cells.

\paragraph{Conclusion:}
 We believe that the physics of complex systems fused with quantum computing will constitute a game changer in modeling and design of large network of RIS devices cooperating in order to transform  real-life propagation environments into a resource for future mobile networks, including SONs and cell-free networks.

\vspace{-0.24cm}

\phantomsection
\section*{Acknowledgment}
The work of GG is supported by the Royal Society Industry Fellowship, grant number INF/R2/192066), and the European Union's Horizon 2020 Industrial leadership project RISE-6G,  No. 101017011.
The work of ZP is supported by U.S. National Science Foundation (NSF) CAREER award, $\#$1750839.


\newpage

\section*{\fontsize{20}{25}\selectfont Metasurfaces for\\ Next-Generation \\Communication Systems} 
\subsection*{Andrea Alu'\textsuperscript{1}*} 
\subsubsection*{\textsuperscript{1}\textit{
		Advanced Science Research Center, City University of New York, USA}} 

\subsubsection*{*\textbf{Corresponding author}: aalu@gc.cuny.edu} 
\vspace{5mm}

\paragraph{Status}
In the quest to maximize data rates and transmission efficiencies in crowded and complex multi-channel environments, reconfigurable intelligent surfaces (RIS) have been raising significant interest in recent years in the field of wireless communications. These concepts have been establishing a powerful platform to control in real time the connections among multiple users and compensate for environment changes \cite{Referen1,Referen2}. In a parallel but for the most part disjoint effort, the electromagnetics community has been developing a new class of engineered surfaces, known as metasurfaces, which can control the local reflection and transmission coefficients point by point at the subwavelength scale in order to demonstrate unprecedented forms of manipulation and control of the impinging wavefront \cite{Referen3}-\cite{Referen5}. It is becoming increasingly evident that these two research areas may largely benefit from closer interactions and joint explorations. On one hand, current research on RIS has been mostly focused on higher-level network and protocol aspects, basing the hardware description on often naïve models, e.g., treating the RIS as a phased array over which the local reflection coefficient can be arbitrarily controlled in space and time, and disregarding physical limitations and engineering challenges at the hardware level. Similarly, metasurface research has often been criticized for being focused on demonstrating fundamental physical principles without an eye on applications and impact on the grand societal challenges of today’s ever-connected world. In the following, we discuss potential opportunities that may arise from connecting these two lines of research, highlighting opportunities that advanced metasurface technology can offer in the context of communications and network systems, and detailing the challenges that need to be addressed to establish a new paradigm for intelligent metasurfaces benefiting wireless networks and communication systems.

\begin{figure}[ht]\centering
\includegraphics[width = .95\linewidth]{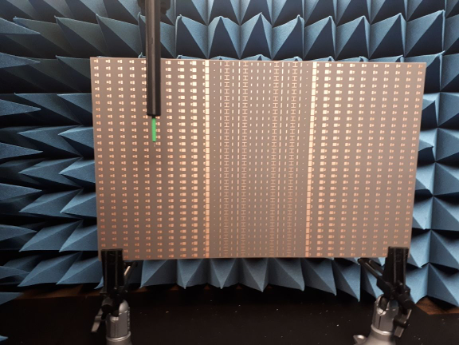}
\caption{\label{fig:overview1} \bf A highly efficient reflective metalens for microwave frequencies, reporting close-to-unity aperture efficiency. (From \cite{Referen8}).}
\end{figure}

\paragraph{Current and Future Challenges}
In discussing the benefits that metasurfaces may offer in the framework of RIS for wireless communication systems, we need to highlight important challenges in their design and implementation. First of all, the subwavelength reactive elements composing a metasurface are prone to suffer from spatial and temporal dispersion: since they need to strongly interact with the local electromagnetic fields, they typically operate close to resonances, and are therefore bound to comply with fundamental principles imposing severe limitations on their temporal response, implying a limited bandwidth of operation \cite{Referen6}. The close interactions of multiple resonators also imply strong coupling between closely spaced elements, inducing spatial dispersion that affects the overall response of the metasurface, which cannot be treated as local. 
In addition to these general challenges, there are limitations stemming from the specific functionality at hand. In the canonical problem of beam steering towards extreme angles, efficiency limitations arise in conventional metasurface approaches that do not take into careful consideration bianisotropy and impedance matching \cite{Referen5}. In \cite{Referen7} we introduced the concept of metagratings, which addresses some of these limitations by enabling beam steering with unitary efficiency even for extreme angles, and lifts the need for dense arrays of resonant elements by combining the concepts of gratings and of complex meta-element designs, hence also alleviating the spatial and temporal dispersion of these elements. Figure \ref{fig:overview1} shows a recently realized metalens based on this approach, showcasing record-high focusing efficiency and numerical aperture for a microwave metasurface \cite{Referen8}. 

Another fundamental limitation of conventional metasurfaces is dictated by time-reversal symmetry and reciprocity: the overall scattering matrix of a conventional RIS obeys strict symmetries when linking different points in its surroundings. Lifting this restriction enables opportunities to better exploit the scarce frequency spectrum, for instance using the same frequency channel to transmit and receive in the context of full-duplex communications \cite{Referen9}. At the same time, breaking reciprocity requires special materials that are difficultly compatible with modern integrated circuits. Alternative opportunities to enable nonreciprocal operations in an integration-friendly platform need to be pursued, for instance leveraging temporal modulation schemes \cite{Referen10,Referen11}. 
Time variations are also important in the context of reconfigurability. To date, most metasurface technology is static, but recent efforts to reconfigure its response using varactors \cite{Referen12}, switches, liquid crystals \cite{Referen13}, graphene \cite{Referen14} or phase change materials \cite{Referen15} have been explored. Important challenges to making them compatible with a highly efficient response is necessary, possibly through the use of active and amplifying elements. Finally, the linearity of conventional metasurfaces poses itself limitations on the overall available responses. Being able to carefully tailor their nonlinear response \cite{Referen16}, we may enable other interesting opportunities, such as ad-hoc frequency transformations and more complex responses. At the same time, these efforts need to be accompanied by careful considerations associated with possible signal distortions and unwanted channel mixing.

\begin{figure}[ht]\centering
\includegraphics[width = .95\linewidth]{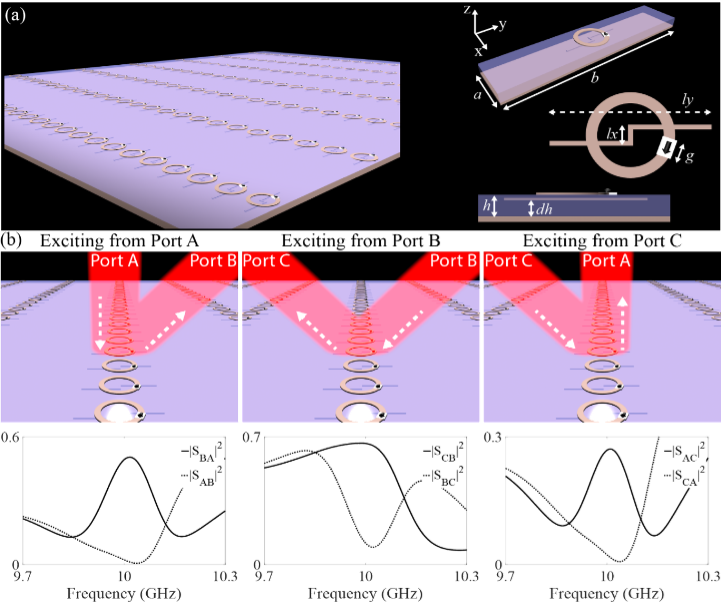}
\caption{\label{fig:overview2} \bf A nonreciprocal metasurface enabling a largely asymmetric scattering matrix for full-duplex operation. (From \cite{Referen19})}
\end{figure}

\paragraph{Advances in Science and Technology to Meet Challenges}
Bringing together the metasurface and RIS communities may tackle these challenges and focus on solutions pushing forward a new platform for reconfigurable, highly intelligent metasurface technology that can revolutionize wireless communication systems. Integrating active elements over the surface may overcome several of the limitations mentioned in the previous section associated with passivity, enhancing transmission efficiency and broadening the bandwidth of operation \cite{Referen17} beyond the limits of passive technology. Temporal modulations can also be exploited in this context, leveraging parametric phenomena to amplify signals and go beyond the bandwidth limitations of passive elements \cite{Referen18}.
Efficient and fast reconfigurability is also an important aspect that needs to be addressed in next-generation metasurfaces for wireless communication systems, ideally embedding opportunities for self-reconfiguration through smart sensors that monitor changes in the environment, e.g., position of the users and channel modifications, and autonomously corrects for these changes. As mentioned above and shown in Fig. \ref{fig:overview2}, controlled time variations can also provide a route for magnet-free 
nonreciprocal responses, ideally suited to enhance the channel capacity and enable full-duplex operations \cite{Referen19}. Parametric mixing in time-varying metasurfaces can also be used to implement frequency transformations, for instance to compensate for Doppler shifts generated by users moving relative to the surface \cite{Referen20}.
Finally, nonlocalities and spatial dispersion, which are nuisance in the conventional operation of metasurfaces, may be engineered using sophisticated designs, enabling the control and manipulation of the impinging signals, and even to impart mathematical operations on them \cite{Referen21}. Leveraging recent advances in network theory, it may be possible to implement ad-hoc signal filtering and processing in space and time of the impinging signals and process them in the analog domain at the speed of light as they interact with our intelligent metasurfaces.

\paragraph{Conclusion}
We envision a bright future in leveraging and bringing together recent advances in metasurface technology and in wireless communications and networks. The strong synergy between these areas holds the promise for largely enhanced communication systems. It will be important to engage the several relevant communities, from the physical layer to the network protocols and algorithms that can enhance the overall system, with the goal of establishing a new class of highly reprogrammable, intelligent, efficient, metasurfaces that can facilitate the next-generation of communication systems.



\newpage

\section*{\fontsize{20}{25}\selectfont Autonomy for Reconfigurable
Intelligent Surfaces! 
A pioneers’ point of view
} 
\subsection*{ Geoffroy Lerosey\textsuperscript{1}* and Mathias Fink\textsuperscript{1,2}} 
\subsubsection*{\textsuperscript{1}\textit{Greenerwave, 6 rue Jean Calvin, 75005 Paris, France}} 
\subsubsection*{\textsuperscript{2}\textit{Institut Langevin, 1 rue Jussieu, 75005 Paris, France}} 
\subsubsection*{*\textbf{Corresp. author}: geoffroy.lerosey@greenerwave.com} 
\vspace{5mm}

\paragraph{Introduction}

It all started back in 2012, when Mathias came back from his countryside house, complaining about his cell phone very poor connectivity there. After days of discussion, inspired by the recent concept of wavefront shaping in the field of optics \cite{Vellekop,Popoff,Mosk}, we had the idea. We would design smart surfaces that can adapt themselves in real time to shape, at will and in a passive way, the propagation of electromagnetic waves. We would hence make the environments electromagnetically smart for greener and more reliable wireless communications. And so we did; Leveraging our knowledge of wave propagation in complex environments, we realized that a tunable mirror where the phase shift is discretized with only two phases is an optimal compromise between hardware complexity and wave control capabilities. We developed our first electronically tunable metasurfaces in the microwave range in 2012 and provided the first laboratory proof of concept in 2013, increasing 10fold the energy received by an antenna in the WIFI frequency band, simply by shaping existing scattered waves. We also filled a first patent application and submitted a subsequent set of seminal publications \cite{Kaina1,Kaina2,Dupre}. 

\begin{figure}[h]\centering
	\includegraphics[width=0.7\linewidth]{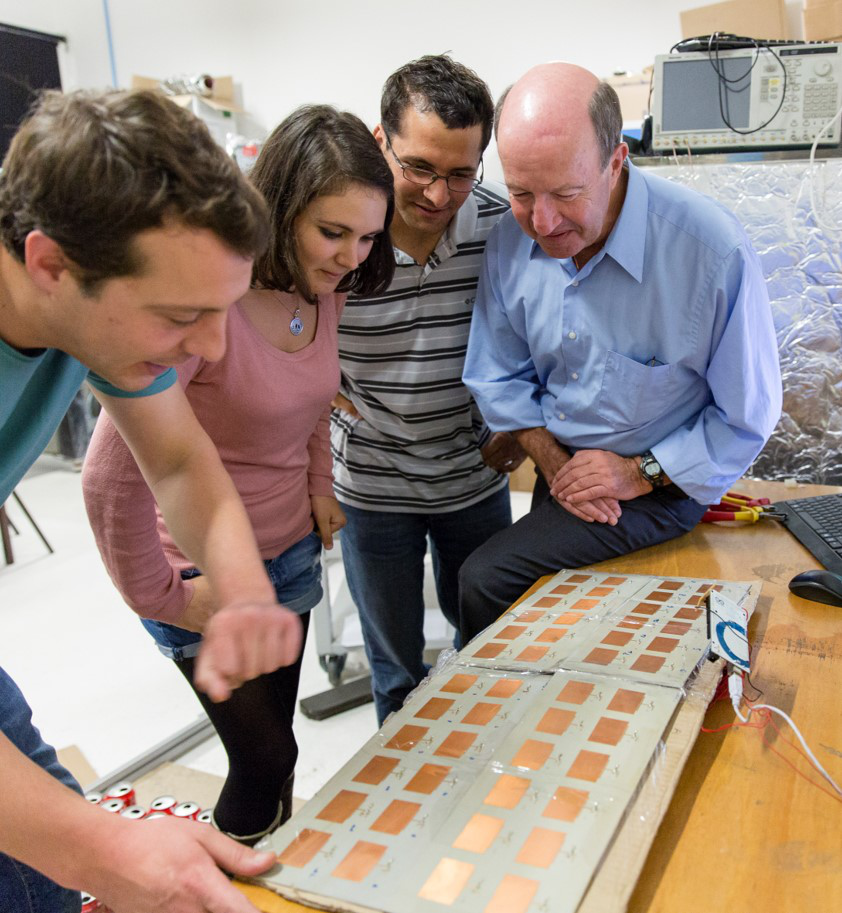}
	\caption{The first tunable metasurface developed at Institut Langevin at 2.45GHz with from right to left Mathias Fink, Matthieu Dupré, Nadège Kaïna and Geoffroy Lerosey (2012).}
	\label{fig1F}
\end{figure}

The work received rather moderate attention in the scientific community for years, except for some teams specialized in metasurfaces or wave physics \cite{Rotter,Chen,Boyarsky}. Yet we received a fantastic enthusiasm when we showcased our concept and provided live demonstrations of passive microwave control using tunable metasurfaces. And this encouraged us to launch a company devoted to commercializing this new technology in the field of wireless communications: Greenerwave. We made tremendous technical progress thanks to the creation of the company and the funding it received, developing industry ready tunable metasurfaces, designing hardware boards to control them, or devising new algorithms to shape waves \cite{Hougne}. Yet the company almost went bankrupt, unable it was to deliver its promises and to push this technology to the market. To survive it had to pivot to other applications based on the same core concept, passive wave control using tunable metasufaces, that are mmWave antennas and radars.

\paragraph{Challenges we faced with RIS}

Of course, our pioneer proposal was much too early for the market, which explains part of this situation. Yet there were other reasons why the company failed to bring reconfigurable intelligent surfaces to the market, of which the two most important ones lie at the very root of the concept. 
On the one hand, making the environment smart for wireless communications in real time supposes to have robust and quantitative information on the wireless links that are established, to be used as feedbacks to pilot the algorithms controlling the metasurfaces (signal to noise ratio, signal to interference ratio, channel state information…). This supposes to have a real time access to deep level proprietary information from chipset manufacturers, either at the device or at the access point level. Greenerwave could never obtain such access, and it failed to deliver convincing proofs of concept of the technology because of this technological barrier. 
On the other hand, a very severe limitation showed up from the business side, where questions of integration very rapidly came into play. Indeed, the technology relies on smart surfaces that need to be distributed and set up in the environment, and the larger the surfaces, the better the performances. This quickly raised concerns about how to power these, how to control these, and finally the need for wires and infrastructures to do so, which clearly limited the interest of potential early adopters.

\paragraph{Future of RIS}
Our long history with RIS therefore urges us to claim autonomy for RIS! We believe that for a massive adoption of the technology, these reconfigurable intelligent surfaces should be as self-standing as possible. And this should in our opinion tackle the two problems we encountered. 
First, of major importance is the ability of the RIS to configure itself for optimal wireless links with minimal feedback from the network, if not none. This would allow any RIS to be compatible with any device and avoid or minimize issues between networks/operators. From a physics point of view it should be possible to add extra hardware on a RIS to allow it to estimate the configuration necessary for optimal beamforming on given users, without any feedback between the users and the RIS. To do so, inspired from holographic concepts, we think that real time optimization of the RIS configuration can be done by a simple intensity measurement on each RIS pixel followed by a smart algorithm, reducing at its minimum the complexity to add on the RIS. A relaxed approach could be to separate RIS in two categories; the first one would consist in infrastructure RIS, close to access points, controlled by the network and used to optimize a global coverage, while the second one, end-user RIS, would be closer to the users and controlled by their devices only. Such differentiations would allow to use the best feedbacks on each side for optimal quality of service.
The second mandatory point for us for RIS to be massively deployed is that they need as minimal infrastructure to operate as possible. This means first the possibility to control them wirelessly and in real-time, whether it is from the network or from the devices. From the base station point of view, such possibility should be granted by 5G, since RIS could become part of the IoT nebulae, and be addressed wirelessly with latencies as low as milliseconds. From the users’ point of view, real time optimization could also be implemented, but how to do so practically remains an open question. In the meantime, RIS should be energy self-sufficient, to enable cable free installation and close to zero maintenance. This requires a very optimal design in terms of energy consumption, which is relatively easy to realize in the GHz range, but more challenging in the mmWave domain. Furthermore, RIS should be able to extract energy from their environment. To do so, solar panels could be used outdoor, while RF energy harvesting could be envisioned indoor, with RIS extracting energy from the electromagnetic waves themselves, akin to RFID.

\begin{figure}[ht]\centering
	\includegraphics[width=0.7\linewidth]{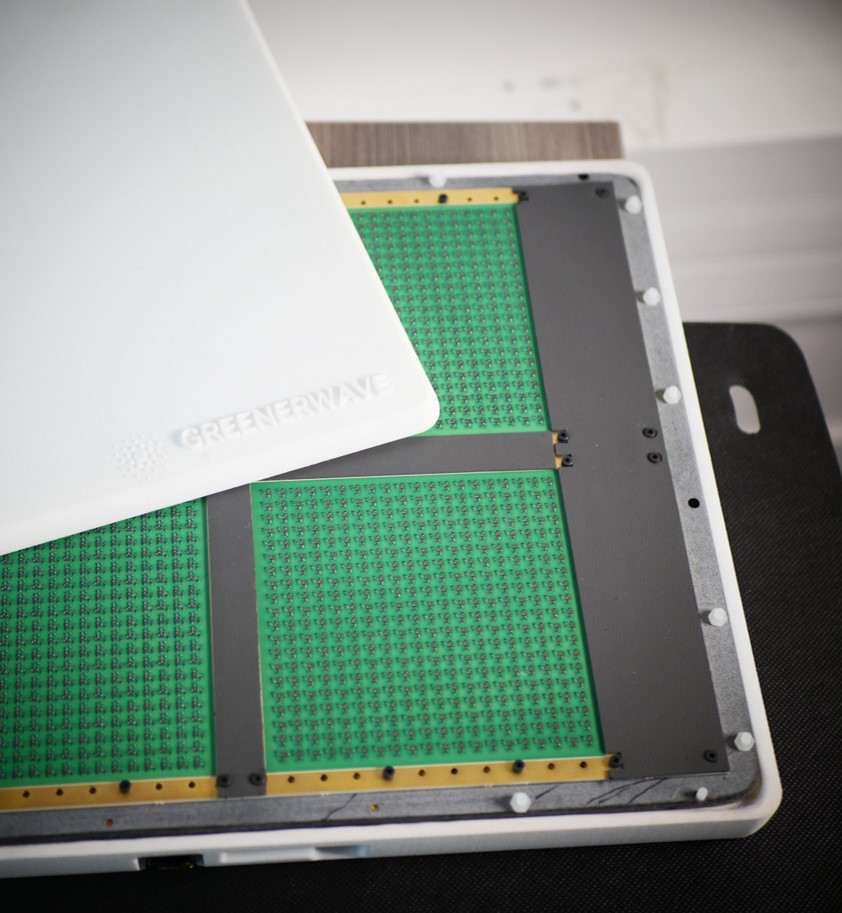}
	\caption{A 20cm*20cm dual polarized tunable metasurface at 26-30GHz, developed by Greenerwave for 5G mmWave access point extender (2021).}
	\label{fig2F}
\end{figure}

\paragraph{Conclusion}
With the enormous interest raised by RIS in the wireless communications community \cite{Renzo1,Basar,Bjornson}, we are extremely excited to see that our pioneering proposal and Greenerwave’s original goal may finally become a reality. There are challenges to tackle to do so but making the environment electromagnetically smart certainly goes in the sense of history, as it is already done for sound, heat, and even light management. Autonomy for RIS would certainly help to go in this direction. Until then, RIS will remain limited to some specific applications such as passive access point extenders for mmWave, a topic we are already working on \cite{GW}.

\phantomsection
\section*{Acknowledgment}
This work has been partly funded by the European Commission through the H2020 project through the RISE-6G, HEXA-X (Grant Agreement no. 101015956).


\newpage

\section*{\fontsize{20}{25}\selectfont Some Perspectives on\protect\\ Space-Time Coding Digital Metasurfaces} 
\subsection*{Vincenzo Galdi\textsuperscript{1}* and Tie Jun Cui\textsuperscript{2}} 
\subsubsection*{\textsuperscript{1}\textit{
		Fields \& Waves Lab, Department of Engineering, University of Sannio, Benevento, Italy}} 
\subsubsection*{\textsuperscript{2}\textit{
		State Key Laboratory of Millimeter Waves, Southeast University, Nanjing, China}} 
\subsubsection*{*\textbf{Corresponding author}: vgaldi@unisannio.it} 
\vspace{5mm}

\begin{figure}\centering
	\includegraphics[width=0.95\linewidth]{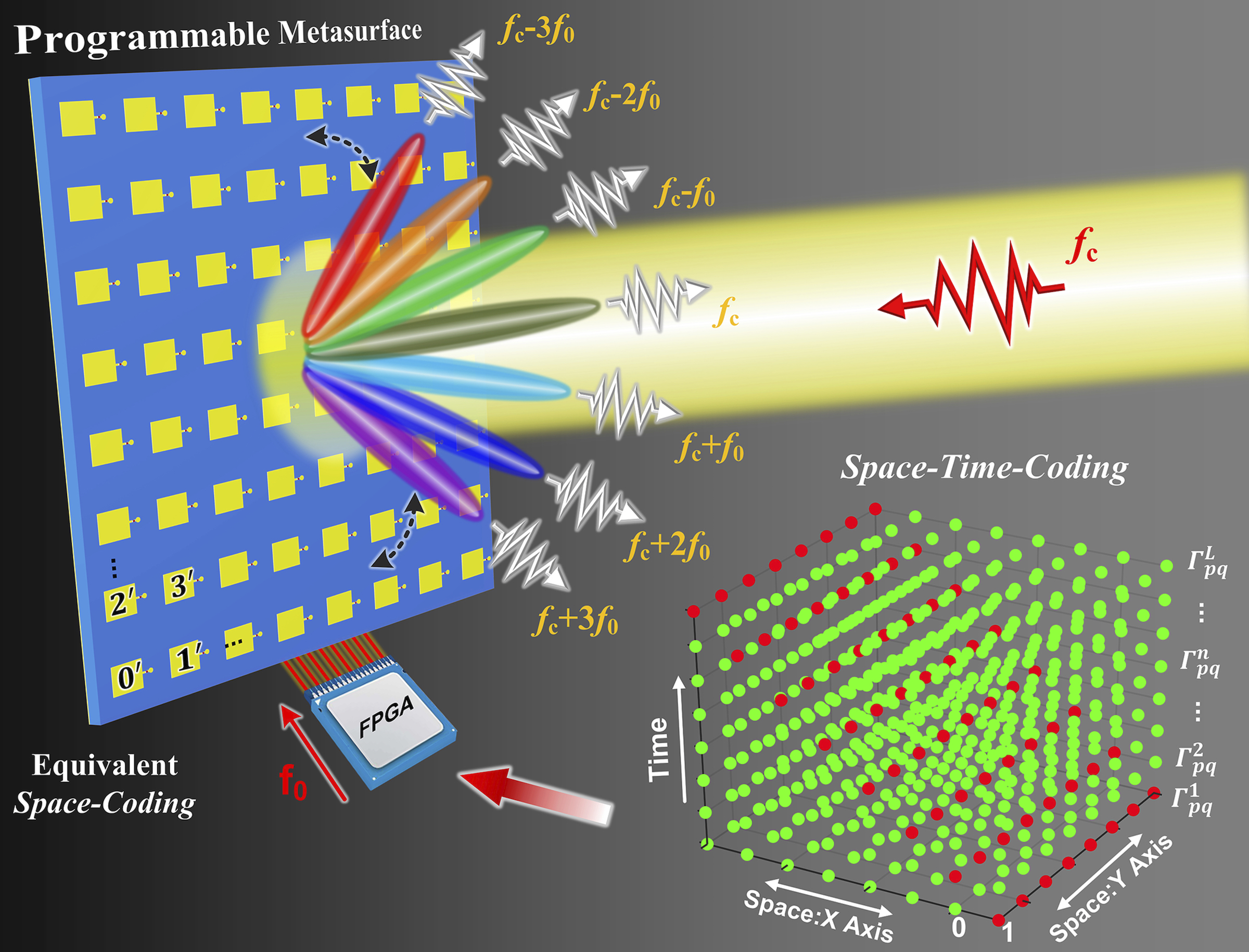}
	\caption{Conceptual illustration of a STC-DM platform (reproduced from \cite{Zhang:2018st}).}
	\label{TJC-VG_Figure1}
\end{figure}

\paragraph{Introduction}
Space-time coding digital metasurfaces (STC-DMs) are a class of recently introduced spatio-temporal metastructures which enable simultaneous field manipulations in both space and frequency domains \cite{Zhang:2018st}. As conceptually illustrated in Fig. \ref{TJC-VG_Figure1}, in their simplest form, STC-DMs essentially rely on a digital, programmable metasurface platform \cite{Cui:2014cm} for which the electromagnetic response of each element (yellow square patches) can be reconfigured between two possible states (e.g., in-phase and out-of-phase reflections, associated with the 0/1 bits)
via a switching element such as a positive-intrinsic-negative (PIN) diode. Therefore, each possible response can be encoded in a sequence of bits that can be controlled and re-programmed via a field-programmable gate array (FPGA). In STC-DMs, this concept is further leveraged by introducing dynamic-modulation aspects inspired by time-modulated arrays \cite{Kummer:1963ul} and phase-switched screens \cite{Chambers:2004tp}, whereby the switching is controlled in space and time according to a suitably designed 3-D coding matrix (red and green dots).

This allows sophisticated spatial/spectral field manipulations, such as the {\em harmonic beam steering} illustrated in Fig. \ref{TJC-VG_Figure1}, where an impinging beam is re-radiated into multiple beams at different frequencies and directions, in a controllable fashion \cite{Zhang:2018st}. In principle, higher-bit programmable metasurfaces can be exploited to reduce the phase quantization error and therefore enable a more precise control, and it should be noted that the equivalent phase distributions at the harmonic frequencies generally exhibit a finer quantization than that of the coding elements \cite{Zhang:2018st}. For instance, a 2-bit programmable metasurface combined with a time-coding approach was demonstrated to provide arbitrary multi-bit and even quasi-continuous programmable phases \cite{Zhang:2020dr}.
More recently, the same principles have been exploited to attain {\em nonreciprocal} reflection effects \cite{Zhang:2019br}, and to address the independent and simultaneous syntheses of prescribed scattering patterns at given harmonic frequencies \cite{Castaldi:2020jm}.
Also worth of mention are some alternative platforms  relying on varactor (instead of PIN) diodes, which allow more flexibility in the time-modulation waveform.
This has been shown to enable powerful capabilities in manipulating the spectral distribution of electromagnetic fields \cite{Zhao:2019pt}, and independently controlling the harmonic amplitudes and phases \cite{Dai:2018ic}.

The joint spatial/spectral field-manipulation capabilities enabled by STC-DMs, and their inherent (re)programmability, can be exploited to directly embed digital information in microwave carrier signals, without the need of conventional radio frequency circuits, paving the way for novel software-defined wireless-communication architectures. For instance, in \cite{Zhao:2019pt}, a 
{\em direct modulation} scheme based  on binary frequency shift keying (BFSK) was demonstrated by exploiting a pair of discrete frequencies to represent the bits ``0'' and ``1''. In 
\cite{Dai:2019wc}, a quadrature phase shift keying (QPSK) modulation scheme was implemented, demonstrating real-time video transmission with 1.6 Mbps data rate. 
Within the overarching framework of ``reconfigurable intelligent surfaces'' \cite{Basar:2019wc}, STC-DM platforms and concepts may find many potential applications
in future wireless communication networks.

\paragraph{Emerging challenges}
Within the emerging paradigm of ``information metastructures'' \cite{Ma:2020im}, the temporal dimensionality granted by STC-DM platforms represents a crucial addition to enable novel self-adaptive, cognitive and possibly nonreciprocal functionalities. However, in order to fully exploit the additional degrees of freedom, significantly more complex design approaches are needed, as the joint multifrequency syntheses are inherently entangled, and brute-force optimization approaches may become computationally unaffordable. 

From the modeling viewpoint, the simple time-domain approach developed in \cite{Zhang:2018st}, based on an adiabatic extension of the frequency-domain physical-optics method in \cite{Cui:2014cm}, seems to accurately capture the basic physics and its predictions are in fair agreement with measurements \cite{Zhang:2018st,Zhang:2019br,Castaldi:2020jm}. However, 
with a view toward integrating STC-DM platforms in complex communication systems, more sophisticated approaches are needed, especially in the modeling of the switching elements.

Another critical challenge is the implementation of faster switching schemes, enabling higher modulation frequencies. The PIN diodes utilized for the proof-of-concept demonstrations at microwave frequencies \cite{Zhang:2018st,Zhang:2019br,Castaldi:2020jm} can reach switching rates up to hundreds of MHz, which are insufficient for operating at THz frequencies of potential interest for next-generation wireless communication systems.

Also of great interest would be to add some polarization-control mechanisms, in order to enable joint manipulations in the entire (spatial, spectral and  polarization) parameter space.

\paragraph{Future developments to satisfy these challenges}
On the modeling side, more accurate approaches need to be pursued, possibly hybridizing full-wave and circuit simulations.
From the design viewpoint, semi-analytical strategies based on more sophisticated temporal coding sequences have been proposed, which enable 
independent multi-order harmonic manipulations \cite{Castaldi:2020jm}; these approaches look potentially promising, although their spectral efficiency needs to be improved. 
The application of artificial-intelligence-powered design approaches is also of great potential interest, and still largely unexplored for STC-DM platforms.

Concerning faster modulation schemes, emerging platforms based on graphene \cite{Rajabalipanah:2020rs}
 and vanadium dioxide \cite{Shabanpour:2020ur} seem potentially promising for THz frequencies, although experimental evidence is still lacking.
On the other hand, polarization control can be in principle attained by means of anisotropic coding elements \cite{Ma:2020ss}. Preliminary results from our ongoing studies indicate the possibility to  attain the simultaneous conversion of polarization and frequency.

\paragraph{Conclusion}
In summary, we have attempted a compact overview on the state-of-the art, challenges and perspectives of
STC-DM platforms. Overall, the outlook looks very promising, with ample room for applications under the broad umbrella of
 information metastructures and, more specifically, to reconfigurable intelligent surfaces for future wireless communication systems. The reader is referred to \cite{Zhang:2020ra} (and reference therein) for further details.



\newpage

\section*{\fontsize{20}{25}\selectfont The Future of Intelligent\\ Wavefront Shaping for Smart Radio Environments} 
\subsection*{Benjamin W. Frazier\textsuperscript{1}* and Steven M. Anlage\textsuperscript{2}} 
\subsubsection*{\textsuperscript{1}\textit{
		Applied Physics Laboratory, Johns Hopkins University, Laurel, MD 20723, USA}} 
\subsubsection*{\textsuperscript{2}\textit{
		Department of Electrical and Computer Engineering, Department of Physics, Quantum Materials Center, University of Maryland, College Park, MD 20742, USA}} 
\subsubsection*{*\textbf{Corresponding author}: Benjamin.Frazier@jhuapl.edu} 
\vspace{5mm}

\paragraph{Introduction}
As the electromagnetic spectrum becomes more congested and the environments in which we need to operate become more complicated, control over the environment itself becomes necessary to ensure the integrity of wireless communication channels. Wavefront shaping with programmable metasurfaces allows wave fields to be manipulated in both time and space, providing a method to interact with the environment. When coupled with deep learning, intelligent wavefront shaping serves as a catalyst, enabling smart radio environments and unlocking applications beyond traditional wireless communication networks. In this paper, we discuss the outlook of intelligent wavefront shaping for wave propagation in complex environments and highlight its transformative potential.

\paragraph{Emerging challenges}
Within the emerging paradigm of ``information metastructures'' \cite{Refer1}, the temporal dimensionality granted by STC-Modern radio frequency (RF) imaging and communications systems operate in complex environments that are susceptible to multipath reflections which scramble propagating electromagnetic waves. Transmitted signals in these environments experience random spatio-temporal fluctuations which degrade or disrupt performance, particularly when combined with competing RF emissions and a congested electromagnetic spectrum. A ``smart'' radio environment (SRE) must be able to handle such conditions, adapting on-the-fly to optimize a metric for the wireless channel \cite{Refer1,Refer2}. The optimization should be performed over the entire propagation path, not only at the endpoints as with traditional wireless systems. 

\begin{figure}[ht]\centering
\includegraphics[width = .95\linewidth]{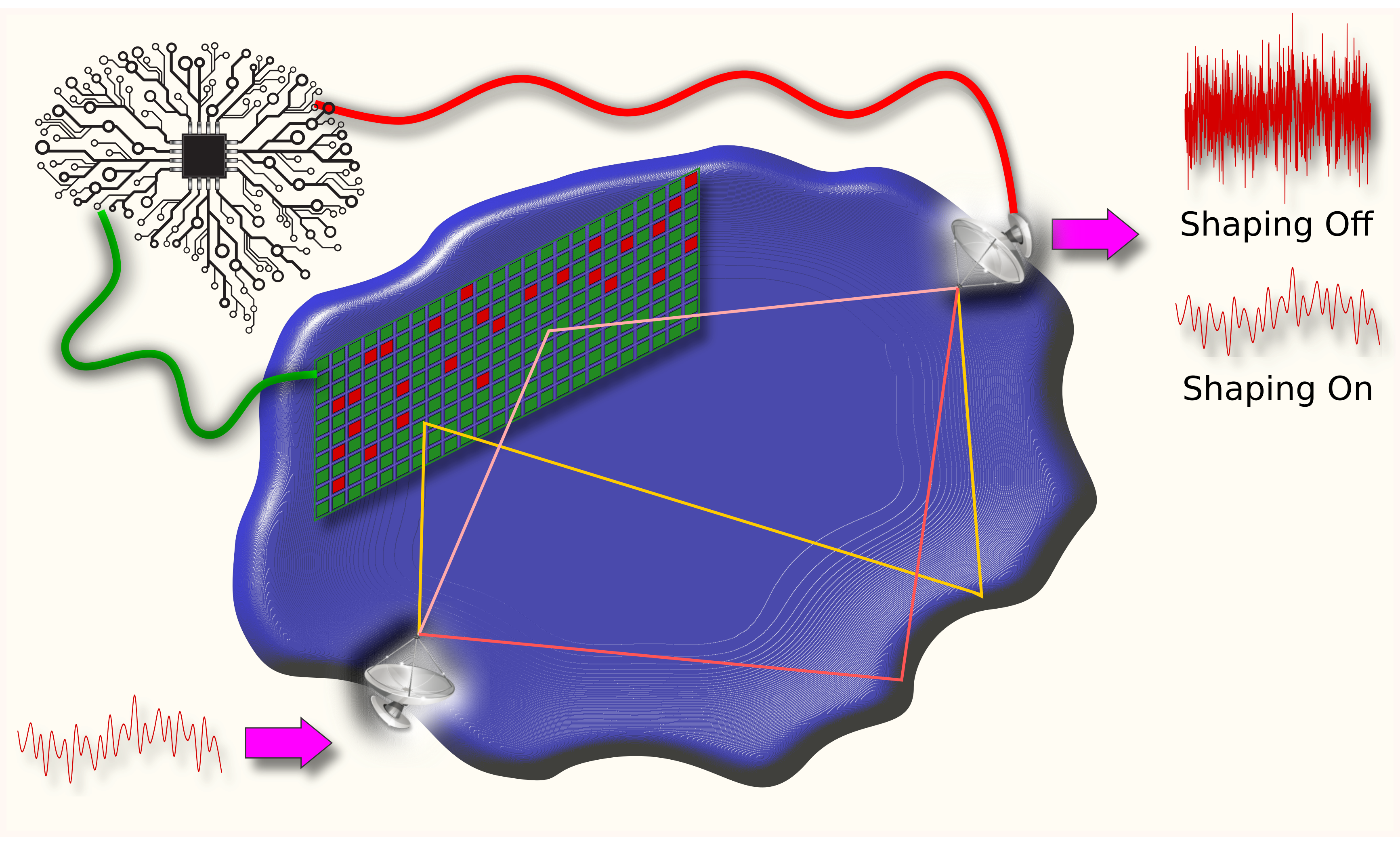}
\caption{\label{fig:overviewA} \bf Conceptual diagram of a smart radio environment (SRE). \normalfont A modulated signal input to a complex environment, such as a wireless network, is scrambled through constructive and destructive interference between the multiple paths. A reconfigurable metasurface is then leveraged to program the environment through intelligent wavefront shaping, reducing the interference and allowing the input signal to be recovered. Deep learning provides the intelligence, updating the metasurface and responding to changes in the environment on-the-fly.}
\end{figure}

The vision of an SRE (Fig. \ref{fig:overviewA}) is a self-adaptive system that can counter scrambling of electromagnetic waves from the complex scattering environment, ensuring operation even under degraded conditions. The ability to program the environment can be achieved through the use of tunable metasurfaces, which can manipulate their local surface impedance to enable on-demand beamforming \cite{Refer3}; these devices are inexpensive and widely available at RF wavelengths, making them ideal for dynamic wavefront shaping applications.

It is natural to consider the wireless channel for an SRE as a long-range, open-world environment that may be a densely packed urban area or a sparsely populated farm. However, an SRE is also valuable in enclosed environments, such as a train station, a passenger compartment on an airplane, or even an office. An SRE therefore has many potential applications, including the ability to enhance 5G communications, protect against electromagnetic interference, induce cold spots at specified locations \cite{Refer4}, realize a microwave cloak around an object \cite{Refer5}, enable computational imaging \cite{Refer6}, and leverage Wi-Fi signals to allow wireless backscatter communications \cite{Refer7}. 

Complex microwave cavities can mimic these larger scale enclosures and are extremely useful as surrogates for prototyping and experimentation \cite{Refer8}. Many metasurface enabled wavefront shaping experiments have been performed in these cavities \cite{Refer9,Refer10,Refer11,Refer12}, demonstrating fine control over the scattering parameters. As such, complex microwave cavities will play a crucial role in the future development of SREs, with intelligent wavefront shaping serving as an enabling technology.

The metasurface is placed inside the scattering medium, intercepting a relatively small number of ray trajectories, sometimes with multiple bounces off the metasurface itself. Therefore, the relationship between metasurface commands and sensed environmental responses is extremely complex, resulting in an ill-posed inverse problem; an issue that is exacerbated when considering multiple distributed surfaces. Central to the vision is intelligence, meaning the device must sense the environment and then deliberately interact with it. The explosion of software defined radios in recent years provides a wealth of powerful and inexpensive hardware to develop sensing prototypes. The distinction of the architecture as ``software defined'' means that general purpose hardware can be reused on many varied applications, significantly reducing cost. 

\paragraph{Future developments to satisfy these challenges}
Wavefront shaping applications to date have relied on brute force trial and error or stochastic search algorithms \cite{Refer11,Refer13}; however, the inherent complexity makes this an ideal place to leverage deep learning. Deep learning has enjoyed great success in the design of metasurfaces, but has had less use in dynamic wavefront shaping applications. To successfully field a deep learning solution, we need to address concerns with long processing times as well as size, weight, and power. Traditional deep learning systems require tremendous amounts of data to train. The acquisition time for sufficient training data may be longer than the coherence time of the environment \cite{Refer2}, resulting in a trained solution that is no longer accurate. The concept of reinforcement learning \cite{Refer14} is at the intersection of deep learning and optimal control, and can assist here. It provides a methodology for adjusting the SRE to environmental changes on-the-fly, and has shown great potential to develop optimal control policies in complex and uncertain environments. To alleviate concerns with overwhelming amounts of data, reinforcement learning can be coupled with transfer learning, where information about previous environments is leveraged to accelerate training \cite{Refer15}.

Processing for deep learning is typically performed on expensive and power hungry graphical processing units, so the footprint in terms of both cost and power may exceed allowable margins. Computational efficiency can be increased by compressing deep learning models through quantization and pruning \cite{Refer16}. The growth of edge intelligence for connected devices in the internet of things (IoT) has produced a demand for smaller deep learning models and cheaper, more efficient processing, which will lead to a wider availability of viable processing platforms.

Cabling and interface requirements grow with the number of unit cells provided by a metasurface. The ability to address individual unit cells and switch states as needed is critical to achieve a practical SRE. Connectors and large cable runs form bottlenecks and tend to be the weakest links in a system, so the capability of addressing unit cells wirelessly without further corrupting the environment is highly desired for large element counts.

\paragraph{Conclusion}
Future wireless systems, including 5G and IoT devices, will increasingly rely on the ability to program the environment as the electromagnetic spectrum becomes even more congested. Intelligent wavefront shaping using metasurfaces coupled with deep learning serves as a path towards realizing an SRE, which will be a revolutionary breakthrough for imaging and communication systems operating in complex environments.


\newpage

\section*{\fontsize{20}{25}\selectfont Unconventional Sources for Smart EM Environments: An Inverse Scattering Vision}

\subsection*{Marco Salucci\textsuperscript{1} and Andrea Massa\textsuperscript{1,2,3}*}
\subsubsection*{\textsuperscript{1}\emph{ELEDIA@UniTN (DISI - University of Trento),
Trento, Italy}}
\subsubsection*{\textsuperscript{2}\emph{ELEDIA Research Center (ELEDIA@UESTC -
University of Electronic Science and Technology of China), Chengdu,
China}}
\subsubsection*{\textsuperscript{2} \emph{ELEDIA Research Center (ELEDIA@TSINGHUA
- Tsinghua University), Beijing, China}}
\subsubsection*{{*}Corresponding author: andrea.massa@unitn.it}

\vspace{5mm}

\paragraph{Introduction}

Future standards beyond the fifth-generation (\emph{5G}) will substantially
change the way a communication system is nowadays conceived and deployed
\cite{Akyildiz2020}-\cite{Chowdhury2020}. The progressive shift
to millimeter-waves and the need for massive access and ubiquitous
wireless coverage with extreme data throughput will pose unprecedented
challenges in the design of next generation systems \cite{Akyildiz2020}\cite{Chowdhury2020}.
Higher capacity and link reliability, lower latency and power consumption
as well as reduced costs and complexity are just few representative
examples of several goals to be addressed in the forthcoming years
by academic and industry researchers with unconventional solutions
\cite{Rocca2016}\cite{Herd2016}. Clearly, the era of designing
base-stations (\emph{BTS}s) in ideal propagation conditions and without
obstacles between the antenna and the mobile users is destined to
end soon. Indeed, the complex scattering environment where the communication
system is deployed cannot be no more regarded as an uncontrollable
impairment to the overall quality of service (\emph{QoS}). Accordingly,
standard line-of-sight (\emph{LOS}) key performance indicators (\emph{KPI}s)
such as gain, half-power beamwidth, and sidelobe level should be discarded
in favour of \emph{QoS} system-level \emph{KPI}s. A first step towards
this direction is the so-called {}``capacity-oriented'' paradigm
\cite{Oliveri2019}, which takes into account the presence of the
environment as another key {}``stakeholder actor'' for determining
the overall \emph{QoS} at the receiver. Following such a recipe the
overall end-to-end capacity is optimized and, as a result, commonly
undesired features such as grating lobes now enable a profitable exploitation
of the multipath propagation scenario \cite{Oliveri2019}. A step
forward is to consider the obstacles between \emph{BTS} and users
as \emph{enabling factors} for building future \emph{smart electromagnetic
environments} (\emph{SEE}s). According to such a vision, the propagation
channel is not only involved in the design process, but it also assumes
a \emph{positive} role in fulfilling demanding system-level performance
requirements \cite{DiRenzo2019b}\cite{Basar2019}.

\paragraph{Emerging challenges}

Many innovative ideas have been recently proposed to realize the dream
of \emph{SEE}s. One promising (yet widely unexplored) solution is
to cover the walls of the buildings with artificially-engineered passive/active
metasurfaces to control the propagation of \emph{EM} waves through
anomalous reflections breaking conventional Snell's laws \cite{DiRenzo2019b}\cite{Basar2019}. 

\noindent However, there are still many challenges to be addressed
such as the study of (\emph{i}) innovative solutions for the synthesis
of {}``smart \emph{BTS}s'' able to \emph{sense} the surrounding
environment and \emph{opportunistically} exploit the arising scattering
phenomena, (\emph{ii}) new approaches for the design of feasible,
implementable, and environment friendly smart skins for \emph{EM}
field manipulation, and (\emph{iii}) cost-effective materials and
manufacturing processes for a large-scale deployment of candidate
technologies including intelligent reflecting surfaces (\emph{RIS}s)
\cite{Basar2019}. Finally, the synthesis of unconventional sources
on the buildings facades unavoidably encounters paramount challenges
including the realization of the desired macro-scale field manipulation
without covering windows, doors, or other {}``forbidden'' regions.
New solutions will be also required to favour a seamless integration
with the architecture of the urban scenario by reducing the overall
visual impact, as well.

\paragraph{Future trends and developments towards \emph{SEE}s}

According to the authors' vision, the roadmap to \emph{SEE}s will
benefit from a suitable exploitation of inverse scattering (\emph{IS})
theory and methodologies to take advantage of the scattering phenomena
in future urban scenarios. %
\begin{figure}
\begin{center}\includegraphics[%
  width=0.65\columnwidth]{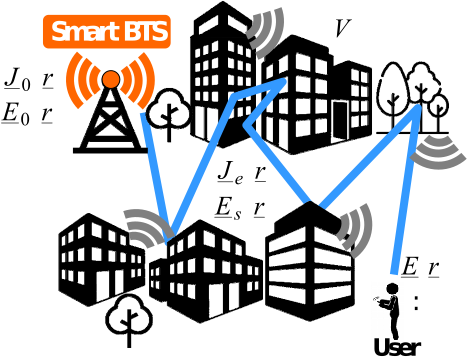}\end{center}

\caption{\emph{SEE} enabled by a {}``smart \emph{BTS}'' opportunistically
exploiting the surrounding environment. \label{fig:opportunistic-sources}}
\end{figure}
According to this paradigm, the \emph{BTS} can be modeled as a primary
source $\underline{J}_{0}\left(\underline{r}\right)$ radiating an
incident field distribution $\underline{E}_{0}\left(\underline{r}\right)$
that induces on the obstacles in a given volume $V$ a distributed
equivalent current $\underline{J}_{e}\left(\underline{r}\right)$,
$\underline{r}\in\mathbb{R}^{3}$ being the position vector (Fig.
\ref{fig:opportunistic-sources}). Thus, the field distribution $\underline{E}\left(\underline{r}\right)$
at the receiver/mobile user terminal is given by the composition of
$\underline{E}_{0}\left(\underline{r}\right)$ with the scattered
field $\underline{E}_{s}\left(\underline{r}\right)$ radiated by $\underline{J}_{e}\left(\underline{r}\right)$
(Fig. \ref{fig:opportunistic-sources})\begin{equation}
\underline{E}\left(\underline{r}\right)=\underline{E}_{0}\left(\underline{r}\right)+\underline{E}_{s}\left(\underline{r}\right)=\underline{E}_{0}\left(\underline{r}\right)+\int_{V}\underline{J}_{e}\left(\underline{r}'\right)\underline{\underline{G}}\left(\underline{r},\,\underline{r}'\right)d\underline{r}'\label{eq:radiation-equation}\end{equation}
where $\underline{\underline{G}}\left(\,.\,\right)$ is the Green's
function. Under such assumptions, it is possible to yield the user-desired
$\underline{E}\left(\underline{r}\right)$ distribution at the receivers
by synthesizing a proper $\underline{J}_{e}\left(\underline{r}\right)$
within $V$. One solution is to design a {}``smart \emph{BTS''}
able to reconfigure $\underline{J}_{0}\left(\underline{r}\right)$
to \emph{opportunistically} exploit the known (deterministically or
statistically) surrounding environment without modifying it (Fig.
\ref{fig:opportunistic-sources}). Accordingly, the excitations $\underline{w}=\left\{ w_{n}\in\mathbb{C};\, n=1,\,...,\, N\right\} $
of $N$ transmit-receive modules (\emph{TRM}s) of the \emph{BTS} become
the degrees-of-freedom (\emph{DOF}s) of the synthesis problem at hand,
while the design is formulated as a global optimization task aimed
at minimizing the mismatch between the synthesized field (\ref{eq:radiation-equation})
and a target distribution $\underline{E}_{t}\left(\underline{r}\right)$
defined over a given observation domain $\Omega$ (Fig. \ref{fig:opportunistic-sources}).
Looking at the same {}``picture'' from a different perspective,
the synthesis of $\underline{J}_{e}\left(\underline{r}\right)$ can
be carried out by designing suitable coatings on {}``smart buildings''
walls or introducing artificially-engineered bio-inspired {}``smart
objects'' in the environment. In this case, the \emph{DoF}s are the
constituent materials and/or micro-level descriptors when printed
sub-wavelength patterns are used to realize unnatural permittivity/permeability
distributions (Fig. \ref{fig:feasible-sources-and-NR}). 

\noindent It is worth pointing out that one paramount advantage coming
from \emph{IS} theory is the \emph{non-uniqueness} of the solution
because of the unavoidable presence in $V$ of non-radiating (\emph{NR})
currents generating a null/non-measurable field outside their support.
Consequently, recent advances in the synthesis of reflectarray antennas
leveraging on the possibility to modify the surface currents with
additional \emph{NR} components still radiating the desired $\underline{E}_{s}\left(\underline{r}\right)$
\cite{Salucci2018} are a viable and effective strategy to realize
\emph{EM} skins fitting geometric/feasibility constraints (e.g., yielding
$\underline{J}_{e}\left(\underline{r}\right)=\underline{0}$ in correspondence
with doors and windows). 

\noindent Last but not least, implementing all mentioned ideas will
also require significant efforts to cope with the computational complexity
arising from the need for accurately modeling electrically-large scenarios.
Towards this end, System-by-Design (\emph{SbD}) approaches are currently
under development \cite{Massa2021}, as well as their integration
with efficient ray-tracing-based forward solvers and customized machine
learning/deep learning strategies to break down the time costs of
the design process \cite{Oliveri2020}. 

\begin{figure}
\begin{center}\includegraphics[%
  width=0.65\columnwidth]{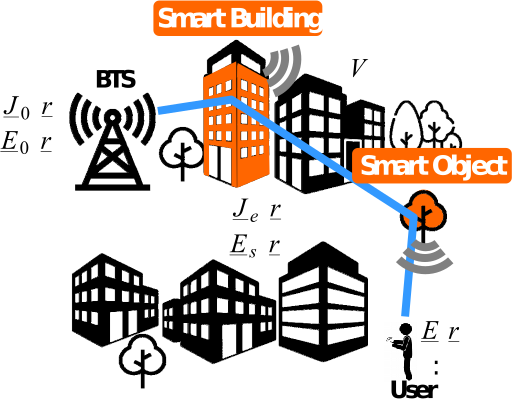}
  \end{center}

\caption{\emph{SEE} enabled by a {}``smart building'' and/or a {}``smart
object'' as feasible sources for \emph{EM} field manipulation. \label{fig:feasible-sources-and-NR}}
\end{figure}

\paragraph{Conclusion}

The envisaged pathway to \emph{SEE}s foresees unprecedented challenges
to be addressed. In this work, new ideas and methodologies for synthesizing
unconventional feasible/opportunistic sources have been illustrated
based on \emph{IS} theory and optimization. 

\phantomsection

\section*{Acknowledgment}
This work benefited from the networking activities
within the Project \char`\"{}CYBER-PHYSICAL ELECTROMAGNETIC VISION:
Context-Aware Electromagnetic Sensing and Smart Reaction (EMvisioning)\char`\"{}
(Grant no. 2017HZJXSZ) funded by the Italian Ministry of Education,
University, and Research within the PRIN2017 Program (CUP: E64I19002530001).

\newpage

\section*{\fontsize{19}{23}\selectfont Metasurface-Based Wireless Communications} 
\subsection*{Qiang Cheng and Tie Jun Cui\textsuperscript*} 
\subsubsection*{\textit{Institute of Electromagnetic Space and State Key Laboratory of Millimeter Waves,  
Southeast University, Nanjing 210096, People’s Republic of China}} 
\subsubsection*{*\textbf{Corresponding author}: tjcui@seu.edu.cn} 
\vspace{5mm}

\begin{figure}[ht]\centering
	\includegraphics[width=0.95\linewidth]{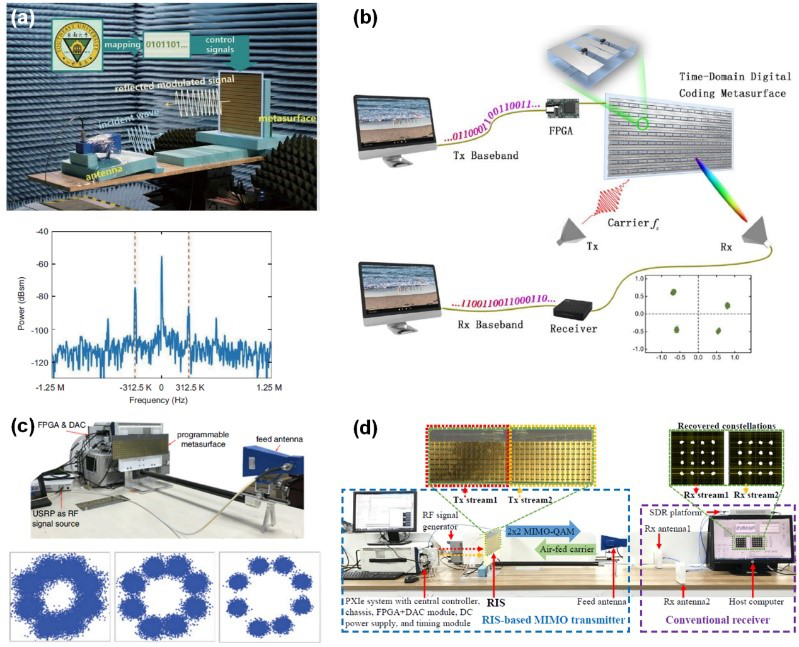}
\caption{Prototypes of the Metasurface-based wireless communication systems, including BFSK (a), QPSK (b), 8PSK (c) and 16QAM (d) systems, where the generated spectrum or constellation diagram are also presented.}
	\label{fig1TJC}
\end{figure}

\paragraph{Introduction}
With the rapid development of wireless communication technologies, explosive growth of personal communications has been witnessed in the past decade, which inspired tremendous efforts to expand the network capacity, reduce the energy consumption, and decrease the system cost. In this paper, we present a brief review and overview on wireless communication systems based on information metasurfaces, which have advantages of simple architecture, low cost, and easy integration. The application of the information metasurfaces for smart propagation environment is also introduced to overcome the multipath fading that is widely encountered in the traditional communication scenarios.

Recent advent of metasurfaces provides new routes for wave manipulations during the wave-matter interactions, and greatly expands the range of available electromagnetic properties of artificial surfaces beyond the natural materials. The metasurfaces are usually made of periodic strongly resonant particles with extraordinary electromagnetic responses. By patterning the particles in a well-defined manner, one can customize the amplitude, phase, polarization, and spectrum of the reflected and/or transmitted waves, thereby offering unprecedented degrees of freedom for device applications such as perfect absorbers, antennas, polarizers, detectors, and sensors \cite{R1}.

Nevertheless, the widespread use of metasurfaces in wireless communications is still plagued by two significant challenges: 1) Dynamic manipulations of the base-station beams are hard to realize by using the traditional passive metasurfaces to boost the signal strength for moving receivers since the meta-atoms and their spatial alignment are fixed; 2) The wave controls are restricted in the physical domain, and digital signal processing technologies cannot be applied during the reflection and/or transmission processes on the metasurfaces.

A feasible solution to overcome these drawbacks is using digital coding and information metasurfaces, which are characterized by digital coding particles (e.g. 0 and 1 with 180$^\circ$ phase difference for 1-bit coding; 00, 01, 10, and 11 with 90$^\circ$ phase difference for 2-bit coding; etc.) \cite{R2,R3}. It was demonstrated that the electromagnetic waves can be manipulated by changing the digital coding sequences. When the digital coding metasurfaces are integrated with field programmable gate array, field programmable metasurfaces are generated, which can be used to control the electromagnetic waves in real time and reprogrammable way. The digital coding particles provide a link between the physical world and digital world, allowing the concepts and signal processing methods in information science to be introduced to the physical metasurfaces, such as Shannon entropy, convolution theorem, and addition theorem \cite{R4}-\cite{R6}. These studies set up the foundation of information metasurfaces, which bridge the physical world and digital world to realize new information systems \cite{R7}.  

By tuning the external biasing voltages of the diodes in each meta-atom, the digital coding metasurface can change amplitude and phase distributions of incoming field dynamically. Additionally, periodic modulation is applied to the surface reflectivity, the baseband information can be directly loaded on the impinging carrier waves without mixer and DAC required by the superheterodyne transmitter, leading to the simplification of the system architecture. 

As examples, several prototypes of metasurface-based communication systems are established to accomplish wireless data transmissions in free space. Various modulation schemes are realized in the experiments, including BFSK, QPSK, 8PSK and 16QAM \cite{R8}-\cite{R12}, as shown in Fig. \ref{fig1TJC}. The measured results confirm the feasibility of direction signal modulations with the time-coding digital metasurfaces, as long as mapping relationships are established between the baseband data and reflection amplitudes and/or phases. Therefore, the metasurface can convert an unmodulated single-tone carrier into a modulated signal owing to the spatial mixing effect. 

\paragraph{Emerging challenges}
One challenge is that the transmission quality greatly suffers from the non-standard constellation diagram, as recognized in Fig. \ref{fig1TJC}, because it highly relies on the dynamic amplitude/phase modulation capability, while the two items are strongly coupled and hard to independently adjust in an arbitrary manner. Hence it remains a formidable challenge for the digital coding metasurface to meet the demand of high-order modulation schemes, such as 256 QAM and 1024 QAM, since the increased constellation points require the reflection amplitudes and phases with extremely high precision for demodulating information correctly.  

The second challenge is the dispersive nature of the metamaterials, which makes it hard to cover multiple operation bands, e.g., the sub-6G and millimeter band simultaneously. In the meantime, the signal modulation on a number of sub-carriers also remains a big trouble, since independent manipulation on hundreds or thousands of harmonics needs large quantization bit number of the reflection/transmission phases for the meta-atoms. 

\paragraph{Future developments to satisfy these challenges}

From the perspective of element design, it is necessary to develop new types of meta-atoms with wide bandwidth, simple structure and low dispersion. To reduce the material loss and control complexity of the digital coding metasurface, the number of diodes should be reduced as much as possible. The biasing networks need be carefully designed to avoid significant impact on the overall scattering performance of the metasurface. 

Additionally, advanced coding theory for the metasurface is expected to play more important roles in signal modulation for multiple harmonics to further improve the communication system capacity. More signal processing algorithms remain to be investigated to enhance the direct data processing ability through the metasurface.     

Toward communication applications, the digital coding and programmable metasurfaces are also critical for controlling and reshaping the wireless channels and environments, reaching the well-known reconfigurable intelligent surfaces (RIS) in the wireless communication community \cite{R13}-\cite{R15}. Such studies will benefit the understanding of the system models and enhance the operation efficiency as well.

\paragraph{Conclusion}
We present a brief review and overview of the metasurface-based wireless communication systems. The mechanism of the direct signal modulations via the information metasurfaces is summarized, and the advantages and limitations of such systems are analyzed in details. The challenges and future research trends are discussed. The metasurface-aided wireless communication brings additional degrees of freedom for wave manipulations and may find important applications in 6G wireless networks.

\phantomsection
\section*{Acknowledgment}
This work is supported by the National Key Research and Development Program of China (2017YFA0700201, 2017 YFA0700202, and 2017YFA0700203), and the National Natural Science Foundation of China (61722106, 61731010).


\newpage

\section*{\fontsize{20}{25}\selectfont AI-enabled RIS-assisted \\
	Wireless Communication} 
\subsection*{ Jinghe Wang* and Shi Jin*} 
\subsubsection*{*\textit{National Mobile Communications Research Laboratory, Southeast University, Nanjing, China}} 
\subsubsection*{*\textbf{Corresponding author}: \{wangjh, jinshi\}@seu.edu.cn} 
\vspace{5mm}

\paragraph{Introduction}
Following the commercialization of the fifth-generation (5G) networks by 2020, researchers are devoted to shaping  the next-generation communication system, namely sixth-generation (6G). However, high complexity networks, high cost hardware, and high energy consumption are becoming crucial issues in 6G. Therefore, it is imperative to explore promising technologies that can be innovative and concise, cost-efficient, and resource-saving. With the revolution in electromagnetic (EM) metamaterials, reconfigurable intelligent surface (RIS) stands out in recent years due to its large capacity, low cost, and low energy consumption properties and unique characteristics of shaping the radio propagation environment.

RIS is a kind of artificial EM surface structure with programmable EM characteristics, which is developed from metamaterial technology \cite{cui2017information}. In recent years, RIS is designed to 
achieve the reconfiguration function since its structure or geometric arrangement of meta-atoms can be reprogrammed by the external control signals. So far, the bottleneck problem in conventional wireless communications is the uncontrollable wireless propagation environment, i.e., signal attenuation usually effects the quality of service (QoS), mutlipath propagation usually leads to various fading, reflection and refraction usually cause interference. Based on RIS, those uncontrollable characteristics are expected to be broken through by tailoring the wireless propagation environment, which can significantly improve the performance of the wireless system.

Over the past several years, considerable researches have been presented. Related research areas of RIS mainly includes two aspects, one is the theoretical research based on mathematical models and the other is the functional implementation and performance measurement based on RIS prototypes. For the first aspect, performance enhancement by RIS phase shift design is one of the most essential topics. For example, reference \cite{wu2019intelligent} firstly proposed joint active beamforming design at the BS and passive beamforming design at RISs to solve the power consumption minimization problem. Reference \cite{han2019large} firstly obtained a closed-form solution of the RIS optimal phase shift design for the RIS-aided massive multiple input single output (MISO) system when only statistical channel state information (CSI) is available. Reference \cite{huang2019reconfigurable} presented two computationally efficient energy efficient maximization algorithms for the BS transmit power allocation and the RIS reflecting elements design. As for the second aspect, the \textit{RFocus} RIS system of MIT \cite{arun2019rfocus}, the RIS prototype of Tsinghua University \cite{dai2020reconfigurable}, and the \textit{ScatterMIMO} of UCSD \cite{dunna2020scattermimo} are prototype verification studies of RIS. Reference \cite{tang2020}\cite{tang2021} present the path loss modeling and measurements for RIS in sub 6G and milimeter-wave frequency band validated by solid numerical measurement results.

\begin{figure}[ht]\centering
	\includegraphics[width=0.3\linewidth,angle =270 ]{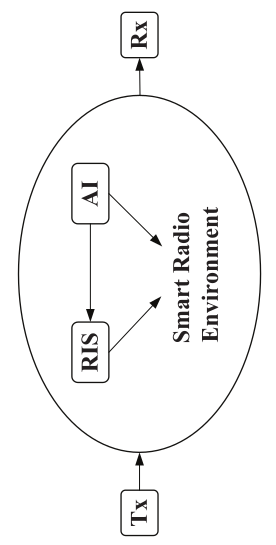}
	\caption{Paradigm of configuring smart radio environment based on AI and RIS.}
	\label{fig1J}
\vspace{-0.3cm}
\end{figure}

\paragraph{AI-enabled RIS-assisted Wireless Communication}
Conventional RIS-assisted wireless communication optimization problems are tackled through various optimization methods based on extracted mathematical model, which include numerous iterations and a large amount of computation. Moreover, some modules in RIS-assisted systems can not be described well by those models. Artificial intelligence (AI), therefore, is of great significance in RIS-assisted systems for processing data without concrete mathematical equation. The paradigm of configuring smart radio environment based on AI and RIS is shown in \textbf{Figure \ref{fig1J}}.

The most widely used technology in AI is machine learning. Machine learning is a kind of algorithm that can automatically analyze and obtain rules and characteristics from massive raw data.Generally speaking, deep learning, reinforcement learning and federate learning are the three paradigms which are extraordinarily effective approaches for designing RISs.

\paragraph{Deep Learning in Signal Detection and Channel Estimation}
The deep learning(DL) approach presents a multi-layer like-brain neural network, which maps the input mass data to the desired output. It is able to mine abstract distribute feature representation to effectively tackle optimization problems. Moreover, these feature patterns can be easily passed to new data, making new data quickly adapt to the environmental changes, which indicates the good generalization ability. 

For those RIS-assisted wireless communication systems,  DL methods are effective for the signal detection and channel estimation problems. Firstly, signal detection at receiver is inherently a classification problem. Conventional detection tasks need the channel estimation, which is challenging in RIS-based radio environment. Therefore, the DL-based detector in the receiver in RIS-assisted wireless communications can be proposed to substitute model-based solutions. Secondly, channel estimation is a regression problem. It is typical that channel estimation at the receiver is carried out by sending pilots from the transmitter.  In RIS-based wireless communication systems, channel estimation is practically challenging due to RISs' passive characteristic. Thus, the DL-based estimation can be proposed to estimate both direct BS-UE channel and cascade BS-RIS and RIS-UE channel.

\paragraph{Reinforcement Learning in RIS Phase Shift Design}
A reinforcement learning (RL) approach utilizes an agent to interact with the environment and learns how to take actions in the next state. Agent learning is to determine the optimal strategy to maximize long-term cumulative rewards. Particularly, deep reinforcement learning (DRL) is the combination of DL and RL, which integrates the strong understanding ability of DL and the decision-making ability of RL.

In RIS-assisted wireless systems, the phase shift design of RIS is crucial to maximize the benefits of deploying RIS for the best system performance. DRL-based beamforming approach can maximize expected performance metrics by appropriately designing the RIS phase shift, which is applied as instant rewards to train the DRL-based algorithm. Therefore, the deep deterministic policy gradient (DDPG) algorithm can be presented in RIS-assisted wireless systems for optimally learning deterministic strategies in high dimensional continuous action space.

\paragraph{Federate Learning in Over-the-air Computation}
The federate learning (FL) approach is an efficient machine learning algorithm between multiple participants or multiple compute nodes. It enables mobile devices to collaboratively learn a shared model without frequent data exchange between mobile devices and servers. Over-the-air Computation (Aircomp) provides a new simultaneous access technology for FL to support fast local model aggregation by taking advantage of the signal superposition characteristics of multiple access channels.

For FL-based Aircomp, minimizing the aggregation error of mean square error (MSE) quantization is the key issue to improve the learning performance. In RIS-assisted FL system, RIS is applied to improve the propagation environment to reduce model aggregation errors and increase the convergence rate of FL, since RIS can control the phase shift to obtain the desired channel response, and the MSE of model aggregation can be reduced with the help of RIS. 

\paragraph{Challenges and Future developments }
Due to tremendous dimensionality, those supervised machine learning schemes heavily relies on the large training data sets to guarantee the system performance. However, standardized data sets for training and testing are finite. Another issue raised when using the model-driven machine learning model is that the offline training period is always large power and resource consuming. For most wireless devices, especially the low-complexity terminals at the edge of the network, the available computing and storage resources are limited. 

Therefore, DRL and FL schemes could be better choices in RIS-assisted wireless networks. DRL learns from experience rather than learns from mass data. FL effectively reduces the huge transmission overhead and meets the requirements of client privacy protection and data security. 
 
\paragraph{Conclusion}
We briefly elaborate AI approaches for RIS-assisted wireless systems for essential problems including signal detection, channel estimation, RIS passive beamforming and Aircomp. Challenges and future developments are attached for guiding the future AI-enabled RIS-aided wireless networks. Considering AI from the initial deployment of RIS will provide us more opportunities to take the full advantages of AI in the performance enhancement of RIS-based smart radio environments.


\newpage

\section*{\fontsize{20}{25}\selectfont Intelligent Surfaces as an Enabling Technology\\ for Holographic Radio} 
\subsection*{Davide Dardari\textsuperscript{1}* and Nicol\'o Decarli\textsuperscript{2}} 
\subsubsection*{\textsuperscript{1}\textit{Department of Electric and Information Engineering (DEI)\\  University of Bologna, Bologna, Italy}} 
\subsubsection*{\textsuperscript{2}\textit{National Research Council (CNR), Bologna, Italy}} 
\subsubsection*{*\textbf{Corresponding author}: davide.dardari@unibo.it} 
\vspace{5mm}

\paragraph{Introduction}

The requirements deriving from the conception of new applications, such as the transmission of holographic videos and autonomous driving, have already put in evidence the limits of the current deployment of the fifth-generation (5G) wireless networks and the and the need to further stress the  performance of the sixth-generation (6G).
Specifically, the extremization of the key performance indexes like data-rate, users' density, reliability, latency and jitter is one of the directions undertaken.
To meet such challenging requirements, technology shifts are necessary in conjunction with the significant increase of the number of antennas and  
 exploitation of higher frequencies.

\paragraph{Emerging challenges}

The higher operating frequency does not come for free. 
The use of millimeter wave and terahertz technologies translates into a larger path-loss, which can be partially compensated by antennas densification and use of large antenna arrays. 
Having massive antenna arrays means also higher hardware and processing complexity and hence higher latency and power consumption that barely scale with the number of antennas. 
Moreover, the shift towards large antennas and high frequency poses new challenges since traditional models based on the assumption of far-field EM propagation fail.
In fact, in classical operating conditions, i.e., small antennas and relatively low frequency, plane wave propagation is assumed (far-field propagation). Conversely, when the antenna becomes electrically large, the operating condition may fall within the Fresnel region (radiating near-field propagation). 
If from one side, the operation in the Fresnel region requires the consideration of new models capable of accounting for this regime, from the other side, it opens new unexplored possibilities to enhance the communication performance through the introduction of 
new design strategies and technologies to exploit it.

\begin{figure}[h!]\centering
	\includegraphics[width=0.7\linewidth]{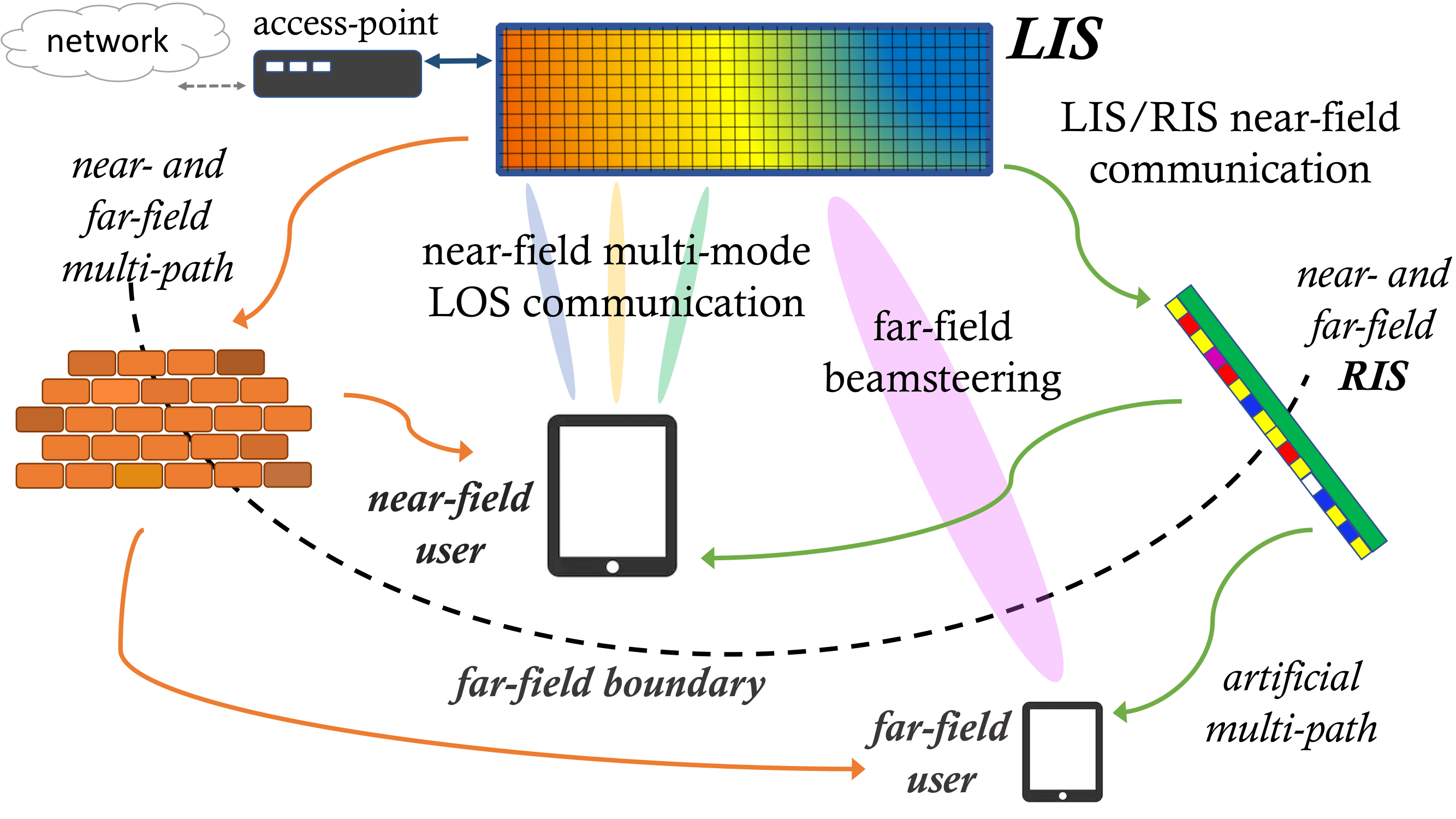}
	\caption{Holographic communication in the near- and far-field of a large intelligent surface used as antenna.}
	\label{fig1D}
\end{figure}

The term \textit{holographic radio} denotes a new paradigm where a wireless system is capable of fully exploiting the characteristics offered by different EM propagation regimes and thus approaching the ultimate limits of the wireless channel \cite{DarDec:20}.
In other words, holographic radio is intended as the possibility to realize the complete control of the EM wave radiated, reflected and/or sensed by an antenna, with unprecedented flexibility.
At the network level, the holographic capability is obtained if the environment is pervaded  by a possibly large number of devices capable of sensing/control the \emph{EM-space} (see Fig. \ref{fig1D}).

One recent candidate technology to enable the holographic radio paradigm is given by metamaterials, which represent the building blocks for realizing \emph{RIS}
\cite{Tre:15}. 
In fact, metamaterials allow the manipulation of the EM field at an unprecedented level of detail, thus enabling the design of specific characteristics in terms of reflection, refraction, absorption, polarization, focusing and steering, when used as reflecting surfaces or large antennas. 
In the last years, the idea of deploying semi-passive reconfigurable reflecting elements in the environment has attracted considerable  attention. Such solutions are able to create additional communication channels between a transmitter and a receiver, thus increasing the coverage and the DoF of wireless communication. Using reflecting surfaces, the wireless channel is not a static entity anymore, whose knowledge is used to optimize transmitters and receivers, but it becomes a partially-tunable element \cite{DiRenzoJSAC:20}.

As active antennas, intelligent surfaces can be exploited to increase the number of the design variables allowing to operate directly at EM level for processing electromagnetic waves \cite{HuRusEdf:18}. When such  antennas are electrically large, namely LIS, the exploitation of the near-field propagation can offer several well-coupled \emph{communication modes} between a transmitter and a receiver, even in LOS, thus enhancing the available communication DoF \cite{Dar:J20}. In particular, it has been shown that the  DoF depends on geometric quantities (normalized to the wavelength), and that novel ad-hoc models are required to characterize the radio link in terms of  DoF and path-loss \cite{Dar:J20,BjoSan:20}. 
Nevertheless, it is still not clear what are the actual fundamental limits in more complex scenarios involving multipath, multiple users, relays and reflecting surfaces, like that shown in Fig. \ref{fig1D}, and how to design practical and low-complexity schemes to achieve such limits.

\paragraph{Future developments to satisfy these challenges}

Reaching the ultimate limits in wireless communication cannot disregard the physical limits of EM propagation. 
The approaching of this limits with practical systems requires solving several theoretical and technological challenges, some of them summarized in the following.

\emph{Metasurface technology and modeling:} 
Metamaterial technology has seen significant progress. Some solutions are already available such as dynamic metasurface antennas, based on waveguide-fed metasurface \cite{ShlAleImaEldSmi:20} and multi-beam antennas. 
Despite that, a significant effort is still needed to augment the configurability of intelligent surfaces in different frequency bands as well as to distill  models capable of  describing properly the characteristics of the surface and being sufficiently abstract to be used in system design. 
    
\emph{Non-stationary channel models:} One peculiarity of electrically large surfaces is that the communication channel may be no longer stationary along the surface and the EM propagation may happen in the near-field condition where the wavefront is spherical. 
    Ad-hoc channel models should be developed and validated to account for such non-stationarity, including non-stationary polarization and the effect of multipath caused by near/far-field random scatterers \cite{WuWanHaaAggAlwAi:15}. 
    
\emph{EM-based signal processing:} 
    In perspective, the flexibility offered by metamaterials can be exploited also as a mean to  shift some functionalities, which are typically performed in the digital domain, directly to EM level with the purpose to tackle complexity and power consumption issues, and  reduce significantly the latency, as the processing would be realized at the speed of light \cite{Sil:14,GuiDar:J21}. 
   
\emph{Holographic radio space awareness:}
Obtaining an enhanced awareness of the radio environment is fundamental for network optimization in terms of resource allocation, interference management, coverage and capacity.
The deployment of intelligent surfaces working at high frequencies allows to construct and keep updated 3D \emph{EM images} of the surrounding radio space describing propagation paths, and radio sources position with high accuracy \cite{SarSaaAlNAlo:20}.  
 For instance, thanks to the RIS-based reflectors, it will be possible to ``look around the corner" and hence obtain an unprecedented level of awareness about the EM environment. 
  
 \emph{Localization:} 
 The position of the surrounding devices  can be inferred through the analysis of the phase profile of the received signal in case the devices are located in the EM near-field region (hologram-based positioning) \cite{GuiDar:J21}. In the EM far-field or non-line-of-sight situations, localization approaches taking advantage of the artificial multipath generated by RIS deployed in the environment can be exploited. 
 
 \emph{Channel state information:} 
 The estimation of the CSI is usually one of the most critical tasks in wireless communication. Moreover, when operating in the near-field, the channel is even more informative thus increasing the associated complexity in estimation \cite{DiRenzoJSAC:20}.
    On the other hand, when moving to higher frequencies, obstacles may completely block the signal, and multipath components become sparse so that communication is mainly enabled by LOS conditions. As a consequence, the CSI is expected to be highly correlated to the geometric configuration of antennas, i.e., the relative position and orientation, so that CSI estimation and localization tasks become intimately linked and can be tackled jointly.     
    
 \emph{Network EM theory of information: } There is the need for a full understanding of the fundamental performance limits as well as the development of practical algorithms for operating with wireless networks composed of multiple users, base stations, scatterers, and RIS under different configurations.



\newpage

\section*{\fontsize{20}{25}\selectfont Metasurfaces for Channel Characterization and Beam-Synthesis} 
\subsection*{Okan Yurduseven\textsuperscript{1}* and Michail Matthaiou\textsuperscript{1}} 
\subsubsection*{\textsuperscript{1}\textit{Centre for Wireless Innovation, Institute of Electronics, Communications and Information Technology, Queen's University Belfast, Belfast, United Kingdom}} 
\subsubsection*{*\textbf{Corresponding author}: okan.yurduseven@qub.ac.uk} 
\vspace{5mm}

\paragraph{Introduction}
Metamaterials are sub-wavelength structures that can synthesize artificial electromagnetic (EM) responses that are not seen in nature. One of their particular applications can be given in the context of metasurfaces. A metasurface is a planar aperture synthesized using an array of metamaterial elements. Metasurfaces have been shown to be a promising candidate across a wide range of applications, from cloaking to EM polarization manipulation \cite{chen2016review}. Recently, a new paradigm, compressive sensing, facilitated by wave-chaotic metasurface apertures has gained significant attention \cite{imani2020review}. An advantage of wave-chaotic, compressive metasurface antennas is that they can radiate highly orthogonal modes controlled by a simple frequency sweep \cite{fromenteze2015computational} or dynamic modulation of the metasurface aperture \cite{sleasman2015dynamic}. Using these wave-chaotic modes, it has been shown that the raster scanning requirement of conventional sensing schemes can be relaxed. Hence, instead of relying on a multi-pixel, point-by-point scan, the metasurface layer can be used to encode the scene information and compress it into a single channel. Such an implementation can significantly reduce the number of data acquisation channels, and hence, simplify the hardware architecture.

Intelligent reflecting surfaces (IRS) can be considered a distinct form of metasurface type apertures. An IRS consists of reflection-based unit cells that can collectively alter the amplitude and phase response of an incoming EM wave to synthesize a desired wavefrom of interest on the IRS aperture. One particular application of IRS type apertures in wireless communication systems is to achieve beam-synthesis, ensuring that a communication link is established between an IRS and an end user. As a result, it is evident that an IRS type of aperture requires a channel characterization to be able to steer the synthesized radiation pattern.  

\begin{figure}[ht]\centering
	\includegraphics[width=0.7\linewidth]{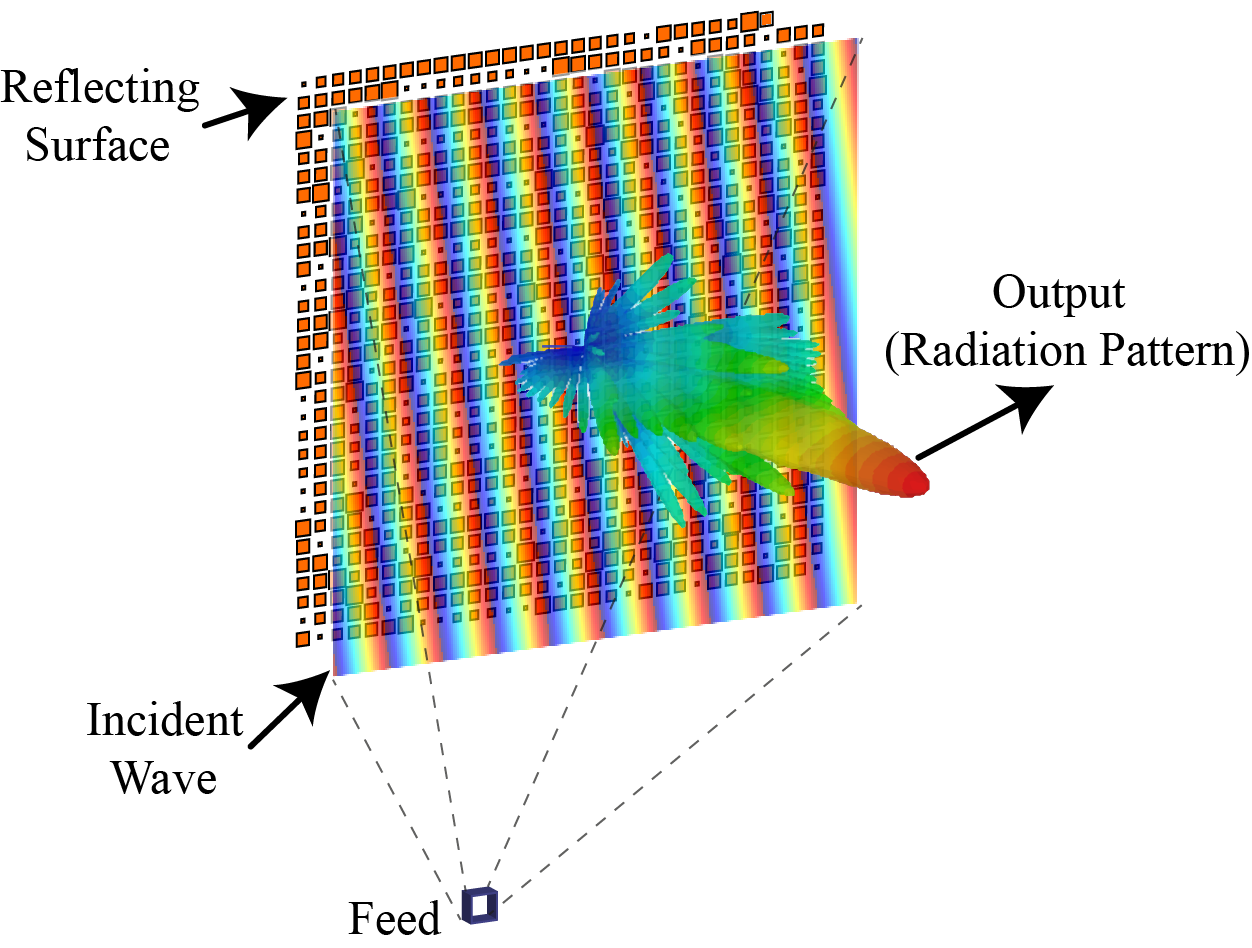}
	\caption{Depiction of a reflective surface for holographic beam-synthesis, wherein the surface converts the incident-wave to a desired radiation pattern.}
	\label{fig1MM}
\end{figure}

\paragraph{Emerging challenges}
An IRS uses a holographic principle to synthesize the desired aperture wavefront as depicted in Fig. \ref{fig1MM}. 

In this holographic framework, the incident wave illuminating the IRS acts as a reference-wave, whereas the desired IRS aperture field distribution that generates the radiation pattern of interest can be considered as the objective function. The role of the IRS is, therefore, to modulate the reference-wave into the objective function when the IRS is illuminated by the reference-wave, similar to an optical hologram. A crucial aspect within this holographic beam-forming framework is that the objective function is required as a-priori information to be able to calculate the IRS surface for a given reference-wave incident. A particularly critical information needed in this process is the direction of arrival (DoA) estimation in order to be able to retrieve the direction of the end user and steer the radiation pattern of the IRS accordingly. 

Conventionally, DoA estimation requires a multi-channel detection unit formed by means of an array topology \cite{vidal2006direction,wang2016non}. Because these techniques require raster scanning on the receiver side, the DoA estimation is achieved by having a full-phase control of each antenna forming the array aperture (phased array technique) or by using antennas mechanically (or electronically) scanned to synthesize an effective aperture. Mechanical scanning can significantly increase the data acquisition time, posing a challenge for real-time operation. On the other hand, phased arrays typically exhibit significant complexity on the hardware layer due to the necessity to have a dedicated phase shifter and power amplifier for each antenna element within the array aperture. Recently, the authors have shown that a single-pixel, compressive metasurface antenna can retrieve the DoA estimation in an all-electronic manner by means of a simple frequency-sweep and using a single channel \cite{yurduseven2019frequency}. One of the significant advantages of this technique is that because it requires a single channel to retrieve the DoA estimation, it can significantly simplify the physical constraints associated with the hardware layer. 

Currently, in wireless communications, IRS apertures are considered as a beam-synthesis technology, leveraging the holographic principle depicted in Fig. \ref{fig1MM}. However, in order to synthesize a radiation pattern of interest using IRS apertures, one needs to know the characteristics of the radiation pattern to be radiated by the IRS. This aspect of the IRS design makes it necessary to have a channel characterization capability built into the IRS design. As a result, an IRS can be used to support multiple capabilities: (a) facilitate compressive sensing to achieve DoA estimation and (b) synthesize a radiation pattern of interest using the retrieved DoA estimation to satisfy the characteristics of the desired channel.   

\paragraph{Future developments to satisfy these challenges}

One of the potential techniques that can be used to address these aforementioned challenges is to use the IRS aperture to achieve both DoA estimation and beam-synthesis. In this context, the IRS can first be designed in such a way that the phase responses of the unit-cells forming the IRS aperture can be modulated randomly to synthesize wave-chaotic bases to facilitate spatio-temporally  varying radiation patterns for compressive DoA estimation. Once the DoA estimation is achieved, the IRS can be switched to the holographic beam-synthesis mode, and using the estimated DoA information as the objective function, the IRS can modulate the reference-wave to steer its radiation pattern in the direction of the end user. 

A significant challenge with this implementation is that both the DoA estimation and the beam-synthesis steps need to be achieved in real-time by the IRS to make sure that the IRS can reconfigure its operation mode and radiation characteristics to capture the dynamic characteristics of the channel that can change over time. This can be achieved using active modulation schemes on the IRS layer facilitated by semiconductor elements with fast switching times. Initial studies have shown that real-time retrieval can be possible using PIN diodes as the switching mechanism \cite{lin2021single}. 

\begin{figure}[ht]\centering
	\includegraphics[width=0.7\linewidth]{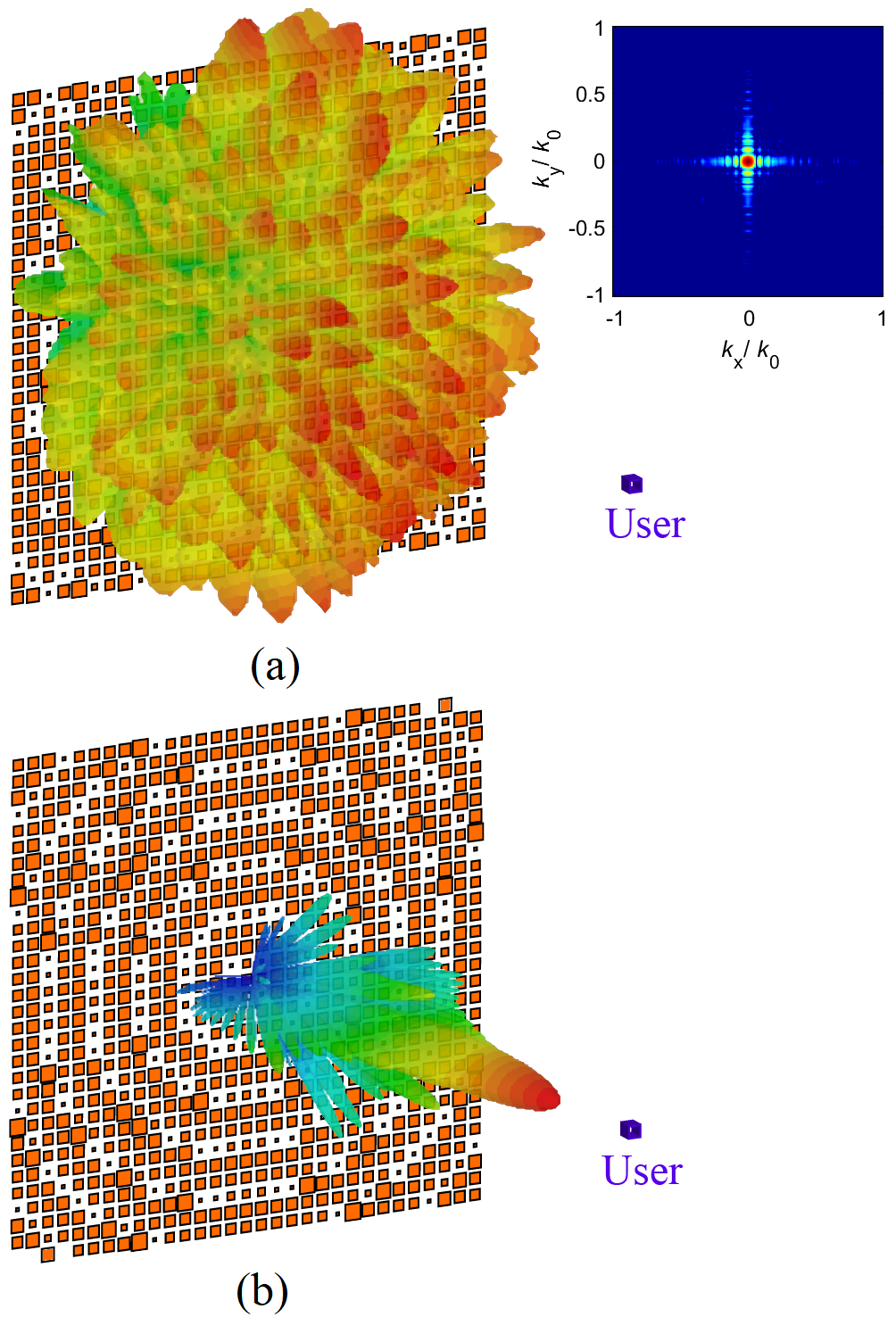}
	\caption{An IRS aperture (a) operating in compressive DoA estimation mode. Retrieved DoA pattern shown as inset (b) operating in beam-synthesis mode.}
	\label{fig2}
\end{figure}

\paragraph{Conclusion}
We have presented a roadmap on the application of metasurface type apertures as IRS structures in wireless communication systems. Leveraging the holographic beam-forming principle, an IRS can be designed to operate in a dual-mode state. By randomizing the complex weights of the unit-cells to facilitate compressive sensing, the IRS can synthesize wave-chaotic radiation patterns that can be used for channel characterization. Once the channel information is retrieved, the IRS can modulate an incident-wave into a desired aperture wavefront, radiating in the direction of the end user using the information provided by the DoA estimation step. Research fronts on IRS types of apertures for beam-synthesis and wave-chaotic metasurface antennas for DoA estimation have recently gained significant traction as two separate tracks. Merging these two techniques into a single design to achieve IRS apertures that can perform real-time channel characterization and beam-synthesis remains as a disruptive technology to be developed in wireless communications.  

\phantomsection
\section*{Acknowledgments}
The work of O. Yurduseven was supported by a research grant from the Leverhulme Trust under the Research Leadership Award RL-2019-019. The work of M. Matthaiou was supported by a research grant from the Department for the Economy Northern Ireland under the U.S.-Ireland R\&D Partnership Programme.



\newpage

\section*{\fontsize{16}{21}\selectfont Ultra-compact `Optical-bench-on-a-chip' Metasurface Toolkits for Ubiquitous Optical Imaging } 
\subsection*{Mitchell Kenney\textsuperscript{1}* and George Gordon\textsuperscript{1}} 
\subsubsection*{\textsuperscript{1}\textit{Department of Electrical and Electronic Engineering,\\  University of Nottingham, Nottingham, UK}} 
\subsubsection*{*\textbf{Corresponding author}: mitchell.kenney@nottingham.ac.uk} 
\vspace{2mm}

\paragraph{Abstract}
We propose a roadmap to widely-deployable multimodality imaging devices that are low-cost and ultra-compact enabled by \emph{metasurface toolkits (MSTs)}, an emerging nanotechnology that can replace bulky traditional glass lenses. We envisage MSTs will play a key role in emerging internet-of-things environments performing tasks from examining road quality, to early diagnosis of disease.

\paragraph{Introduction}
Optical technology is becoming ever more necessary in many walks-of-life, with recent surveys estimating the multimodal imaging market worth £2.25Bn by 2024 \cite{Multi2019}. A particular emphasis also lies upon its translation to smaller more portable devices whilst still encapsulating the smart multifunctionality as full-sized systems. 

Smart-phones have heralded a revolution in hand-held optical imaging devices -- over 7 billion CMOS image sensors are shipped worldwide each year \cite{Insights2018}. -- with applications from conducting biochemical assays to check for disease \cite{Kassianos2015} to identifying damage to roads \cite{Maeda2018} to low-cost widely-deployable screening of water for dangerous pathogens \cite{waterscope}. 
With the rapidly expanding `internet of things' (IoT), devices with sensing capabilities are increasingly integrated into countless environments. To achieve their full potential, image sensors need to offer imaging modalities other than the traditional RGB intensity imaging offered by smartphones. Techniques such as fluorescence,  multispectral, and polarisation-imaging showcase important additional information about the physical world \cite{Rubin2019}, but current approaches require bulky optics made using centuries-old glass technology, akin to large bench-top microscopes, which inhibits down-scaling of advanced imaging systems for ubiquitous IoT devices.

\emph{Metasurfaces} (MSs) are intelligently patterned ultra-thin devices consisting of optically-interacting nanostructures that control light in novel ways \cite{Lee2020}. Metasurfaces exhibit enormous multifunctionality potential in replacing not only conventional optical components but whole optical systems.

\begin{figure}[]\centering 
\subfigure[]{\includegraphics[width=0.5\linewidth]{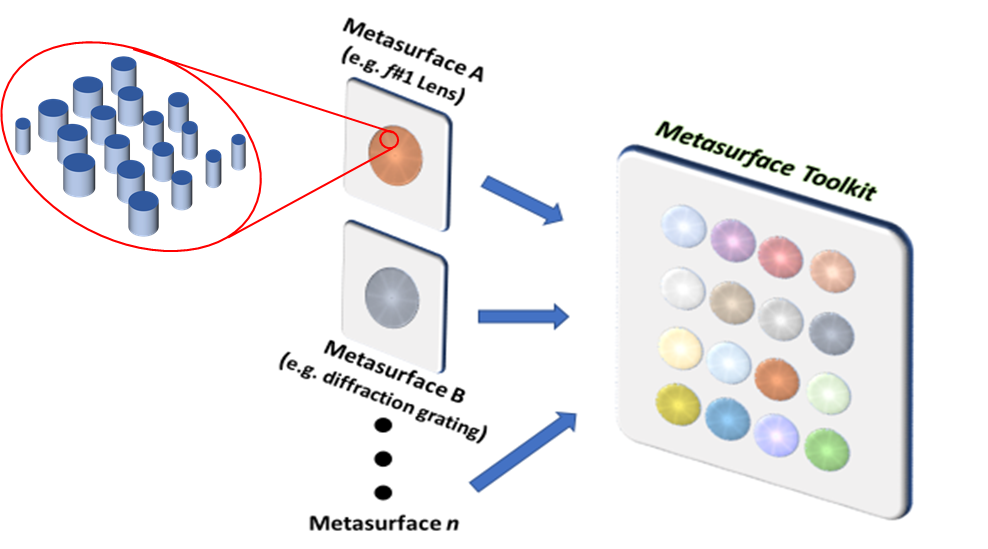}
\label{subfig:singleLayerDiagram}}
\subfigure[]{
	\includegraphics[width=0.3\linewidth]{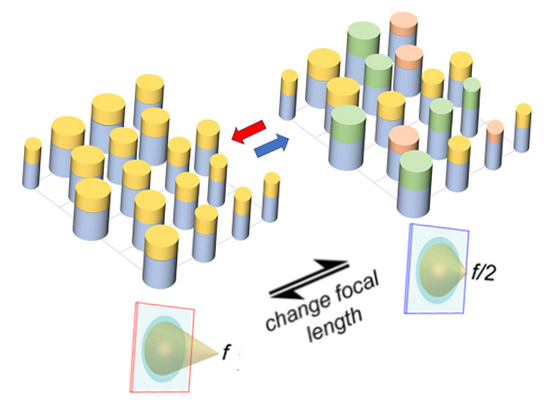}
	\label{subfig:DynamicDiagram}}
\caption{(a) Schematic of individual metasurface devices (Inset: magnified region showing individual nanostructures) being combined for a Metasurface Toolkit (MST). A coin sized MST as a standalone ``Optical-bench-on-a-chip'' consists of an assortment of both simple (e.g. lens, polariser) and complex (e.g. optical-tweezer, vortex phase-plate) optical components. (b) Dynamic MSTs which utilise hybrid nanotechnology. Examples of applications include bi-focal lenses, with two focal lengths ($f$ and $f/2$) being switched. 
	}
	\vspace*{-3ex}
	\label{fig:roadmap}
\end{figure}

We propose a roadmap to widely-deployable advanced imaging devices, based around an emerging concept: compact Metasurface Toolkits (MSTs).  MSTs are coin-sized `checkerboard' samples composed of powerful bespoke metasurfaces, each offering a different functionality or imaging modality that would conventionally require a sizeable optical bench set-up, thus creating a versatile `optical-bench-on-a-chip' (shown in Figure \ref{fig:roadmap}). 
Such devices can be produced at low-cost through high-volume industrial fabrication processes such as nano-imprint or stepper lithography \cite{She2018}.  Further, recent advances in polymer chemistry have enabled dynamically reconfigurable metasurfaces that, when exposed to external stimuli (e.g. temperature, voltage), present different optical behaviours. The imaging functionality can be further increased by stacking multiple MSTs, creating powerful yet ultra-compact systems.

These devices promise transformative applications where advanced imaging is needed but resources and space are constrained, making them ideal for devices as part of the burgeoning IoT. 
Additionally, the customisability of MSs bring applications in fundamental optical research, quantum and communication technologies, and point-of-care healthcare.

\paragraph{Emerging challenges}
There is evidently a need for imaging technology to continue reducing in size while retaining advanced functionality.  However, most optical components are currently manufactured from bulky glasses, which presents a major roadblock. The size of glass-based systems becomes problematic when multiple components are combined together to implement advance imaging modalities, typically requiring bench-sized setups. Additionally, glass components exhibit limited functionalities and are often prohibitively expensive. Because of this, the translation of imaging technologies from large-scale benchtop systems to hand-held portable devices is severely hindered.

Efforts in miniaturizing optical technologies have attempted to address these issues, resulting in well-known devices such as Fresnel lenses and Gradient Index (GRIN) components. Whilst achieving some successes, these devices still lack functionality and require difficult 3D fabrication steps. 

Metasurfaces are an exciting area of optics research, which combine advances in nanofabrication along with bespoke design through tailoring optical properties of nanostructures (e.g. to control amplitude, phase, polarization). 

Although growing interest in metasurface technology has produced impressive results \cite{Chen2012a,Zheng2015a,Chen2018,Khorasaninejad2017}, success in translating these to real-world applications has been hindered due to three key hurdles -- poor-efficiency; difficulty in fabrication and up-scaling to large-area devices; and functionality limited to single-operation devices (e.g. lenses, polarisers).

There have been great advances in metasurface technology for imaging applications, many of which rely on recent developments in using dielectric nanostructures for high efficiency transmissive devices. Only recently have fabrication and computational capabilities advanced sufficiently to produce nanostructures over large areas. Coupled with improved understanding of metasurface performance, this has led to metasurface device efficiencies that rival conventional optical devices. Now, through careful design using nanoscale fabrication, numerous bespoke metasurfaces can be combined onto a single sample -- resulting in the proposed MSTs. 

In the simplest case, different areas of the MST can be addressed to implement different imaging modalities, such as  wide-field, super-resolution, and computational imaging \cite{Lee2020}.  Such designs can also be produced at low-cost due to their compatibility with high-volume industrial fabrication processes such as nanoimprint lithography \cite{She2018}. The imaging functionality can be further increased by producing 3D stacks of MSTs, creating powerful systems that allow numerous optical processing steps in an ultra-compact device \cite{Arbabi2016c,Zhou2018,Lin2019a,Abdollahramezani2020}.

\paragraph{Future developments to satisfy these challenges}
To overcome the challenges of large systems composed of numerous bulky components, we propose Metasurface Toolkits composed of checkerboards of various analogues to conventional optical components, as well as bespoke multifunctional devices that can not be achieved by any singular or combination of conventional optics. Examples of these include bifocal lenses for imaging at different depths, off-axis reflective or transmissive holograms, or even polarisation-sensitive vortex lenses for Optical trapping.

These can miniaturise a whole optical system down to the size of a postage stamp, functioning as an ``Optical-bench-on-a-chip''. These modular devices can be stacked, to produce further enhanced imaging and sensing modalities, such as projection of holograms, through intelligent phase engineering.

Future developments we envision are using deep-learning enabled inverse design approaches to design such stacked Multilayer MSTs (MMSTs) that perform complex functions \cite{Liu2018c}.
These designs could themselves implement passive neural networks capable of light-field processing/sorting. MMSTs will address many issue of single-layer devices, such as aberration correction and 3D imaging. This future direction will combine advances in computer science, optical technologies and modern fabrication techniques, representing a a highly inter-disciplinary approach.
Another promising future direction exploits recent advances in nano-fabrication and polymer chemistry to enable dynamically reconfigurable metasurfaces and MSTs that, when exposed to external stimuli (e.g. temperature, voltage), present different optical behaviours \cite{Hong2004a,Choi2013}. This further increases the range of optical imaging techniques that can performed by an ultra-compact device. 

Many of the approaches in this future-thinking development can be adapted to the existing supply chain for fabrication; namely the micro-electronics industry, which already carry out nanoscale fabrication \emph{en masse} using techniques such as master-stamp lithography or extreme-UV (EUV) lithography. Therefore, they are well suited to enable future scaling up of these advanced device designs.

\paragraph{Conclusion}
We propose the development of Metasurface Toolkits - bespoke postage stamp sized devices composed of nanotechnology-driven metasurface technology, where each metasurface is, by itself, capable of novel and powerful imaging modalities, and when stacked together will unlock even more complex and yet to be realised imaging systems - rivalling standard benchtop systems - resulting in an "`optical-bench-on-a-chip"'. Through intelligent design using deep learning, these systems can be optimised for exciting new science and applications. Finally, the hybridisation of these MSTs with polymer technology will result in dynamic devices that respond to changes in external stimuli (temperature, voltage) leading to switchable operation. These MSTs have potential applications in IoT systems, where multimodality and compactness is extremely desirable.


\newpage

\section*{\fontsize{20}{25}\selectfont Metasurface Based Wireless Localization} 
\subsection*{ Orestis Georgiou\textsuperscript{1}* and Cam Ly Nguyen\textsuperscript{2}} 
\subsubsection*{\textsuperscript{1}\textit{Department of Electrical and Computer Engineering, University of Cyprus, Nicosia, Cyprus}} 
\subsubsection*{\textsuperscript{2}\textit{Wireless System Laboratory, Corporate Research
\& Development Center, Toshiba Corporation, Kawasaki, Japan}} 
\subsubsection*{*\textbf{Corresponding author}: georgiou.orestis@ucy.ac.cy} 
\vspace{5mm}

\paragraph{Introduction}

While metasurfaces are posed to revolutionize wireless communications, their potential use for wireless localization has received very limited attention to date.

Traditional localization is usually achieved via a Global Positioning System (GPS). 
This however introduces additional production costs and power requirements to mobile devices, typically consuming about 30mA at 3.3V. 
Moreover, GPS is not accurate indoors thus failing to meet the Federal Communications Commission's (FCC's) mandate requiring network operators to locate those calling 911 to within certain accuracy requirements (50m horizontally and $\pm$3m vertically).
This, together with the need for position-related services, such as, logistics, smart factories, smart cities, autonomous vessels, vehicles, and localized sensing has seen a wealth of other wireless localization methods being developed and widely deployed. 
These are broadly classified as database, angle-based, range-based, and range-free methods \cite{xiao2020overview}. 

\begin{figure}[t]\centering
\includegraphics[width=0.85\linewidth]{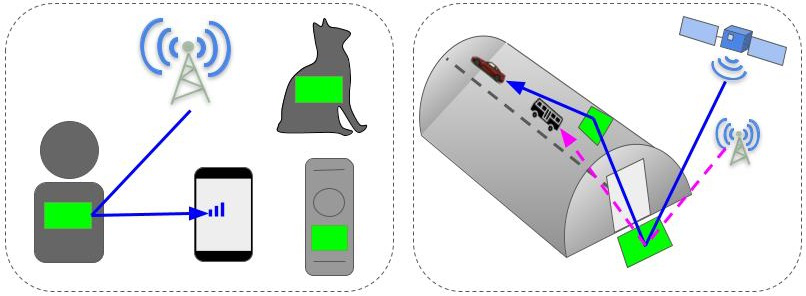}
\caption{Further applications of metasurface based indoor (left figure) and underground localization (right figure).}
\label{fig1OG}
\end{figure}

Recent advances in metasurface design are promising to inject many new capabilities into next-generation wireless communication systems (6G) including enhanced wireless localization and positioning both in and outdoors while also also unlocking novel sensing applications (e.g., Fig. \ref{fig1OG}) \cite{di2020smart}.
This promise has been enabled through the remarkable ability of reconfigurable intelligent surfaces (RIS), namely, to manipulate the propagation of incident electromagnetic waves in a robust and programmable manner thus transforming the wireless channel into a controllable system block that can be engineered and dynamically optimized; the result of many years of R\&D in physics, optics, and material science domains, culminating with the integration of control circuitry into RIS design and unlocking the smart radio environment (SRE) vision.
In the SRE vision, a swarm of low-profile, low-cost, and low-power RIS are densely deployed in- and outdoors and can collaborate and support existing telecommunications infrastructure to: suppressing interference, overcome non-line-of-sight blockage, enhance coverage, improve multi-user throughput and enhance wireless power transfer.

All of these challenges and opportunities are still in their early days and are discussed in other sections of the present roadmap article.
But how does one exploit the added degrees of freedom afforded by dense RIS deployments for improved wireless localization of people and devices?

An initial answer to the above question was attempted by several authors in a point-to-point communication setting (i.e., a transmitter and receiver pair) mediated by a RIS that applies phase differences to the composite wireless channel, the performance gains of which can be captured through the Cramer-Rao lower bound (CRLB) \cite{hu2018beyond,wymeersc2020beyond}.
The CRLB represents the minimum variance of
the error associated with an unbiased estimator, and can be obtained through the inverse of the Fisher information matrix. 
Mathematical and numerical analysis of the CRLB for wireless localization and orientation estimation accuracy suggest considerable accuracy improvements with increasing RIS antenna element count, RIS physical size, and multi-RIS spatial distribution, as one would probably expect.
Multiple trade-offs have also been identified between the different deployment strategies, but a clear winner remains unresolved; further analysis is needed.

Other intricacies of similar RIS-aided localization setups have also been considered, such as the possibility of RIS elements having discrete phase shifts and finite amplitudes \cite{alegria2019cramer}, or the possibility of jointly optimizing communication gains (e.g., improved data-rate) and localization capabilities \cite{he2020adaptive}.
It was concluded that phase resolution is more important to localization accuracy than amplitude sampling density, while adaptive phase shifters  could be efficiently designed and implemented based on hierarchical codebooks and feedback from the terminal.
Authors have very recently also proposed joint positioning, synchronizing, and beam training algorithms, thus capitalizing on RIS ability to focus and improve SNR while employing low-complexity iterative maximum likelihood estimators thus avoiding exhaustive computational search and reducing signaling overheads \cite{wang2020joint}.


\paragraph{Emerging challenges}

Most research on RIS-aided wireless localization has assumed that the underlying channel is largely geometric, comprising of paths of rays that connect a wireless source, the RIS elements, and the terminal while ignoring any rays bouncing off other environmental objects, and also ignoring any mutual EM coupling between RIS elements.
\textbf{Better channel models} are therefore urgently needed both in the far and near field.
Moreover, most proposed approaches, while indeed concluding that significant gains are possible thanks to the RIS focusing abilities and large element count, they do not fully exploit the added spatial and temporal degrees of freedom afforded by the RIS in controlling the proximal electromagnetic field.
\textbf{Non-standard localization methods} should therefore be explored.
Finally, to go beyond algorithmic optimization, more accurate models, and novel localization methods, we must begin to \textbf{experiment with RIS prototypes} in controlled test-beds and practical localization settings in order to truly evaluate performance, discover any possible limitations or novel business opportunities.

\paragraph{Future developments to satisfy the challenges}

A promising approach towards better channel models is already underway by formulating equivalent end-to-end channels in terms of impedance voltages at transmitter/receiver ports which are propagated by electromagnetic fields and mutually coupled among the sub-wavelength unit cells of the RIS \cite{gradoni2020end}.
Building upon such efforts and embedding such models into engineering analysis and optimization is expected to deliver great insights to the field.
Further, it will enable investigations into RIS geometry and periodicity design and their effect on localization performance. 
Channel models could also be extended to capture near-field effects such as wave-curvature as well as uncertainties and peculiarities of the electronics used to control the RIS, e.g., hardware impairments  and frequency/polarization dependencies. 

A different and highly promising approach leverages the RIS ability to alter the radio environment in a controlled and predictable manner.
Coupled with dynamic large databases and machine learning, novel fingerprinting localization solutions have been proposed to exploit the spatial and temporal variability introduced by RIS not only at a single terminal but to the whole deployment region \cite{zhang2020metaradar,nguyen2020reconfigurable}.
Benefits in localization accuracy and coverage using just a single RIS are comparable to what was previously possible through the costly installation of multiple additional access points and antennas.

Finally, metasurface based localization can unlock novel applications that traditional localization cannot (see Fig. \ref{fig1OG}). 
For example, we envision conformal, wearable, and even implantable RIS affecting the electromagnetic properties of its host. 
Further, the dense deployment of metasurfaces can enable localization coverage expansions, e.g, in GPS-unavailable environments such as inside tunnels and buildings while assisted by existing wireless infrastructure.

\paragraph{Conclusion} 
The past decade has witnessed a fast development of reconfigurable metasurfaces, recently posed by wireless engineers as a key technology for next generation communication systems (6G). 
The high focusing capabilities of RISs, their ability to alter/encode waveforms onto impinging signals as well as their unique capacity of producing  differential spatiotemporal radio maps can be capitalized for finely estimating the location of mobile terminals and devices.
To that end, we envision that in the next few years we will see more accurate yet tractable channel models, new system architectures, methods and localization algorithms exploiting novel RIS waveform designs, generative spatial diversity, and fast temporal modulations for improved localization and sensing applications. 
Along with a wealth of theoretical and numerical results, we also hope to see prototypes and test-beds being deployed, exploring both the fundamental and practical capabilities while utilizing new conformal, wearable and even implantable RIS designs, thus enabling novel localization applications and use cases.

\phantomsection
\section*{Acknowledgment}
The authors would like to acknowledge funding from the EUs H2020 research and innovation programme under the Marie Skłodowska-Curie project NEWSENs, No 787180; and Vingroup Joint Stock Company (Vingroup JSC), Vingroup and  Vingroup Innovation Foundation, No VINIF.2020.DA09.


\newpage

\section*{\fontsize{20}{25}\selectfont  Sensing the Environment by Surface Wave Based \\ Metasurfaces} 
\subsection*{Enrica Martini\textsuperscript{1} and Stefano Maci\textsuperscript{1}*} 
\subsubsection*{\textsuperscript{1}\textit{Department of Information Engineering and Mathematics, University of Siena, Siena, Italy}} 
\subsubsection*{*\textbf{Corresponding author}: macis@dii.unisi.it} 
\vspace{5mm}

The new generation of connectivity systems based on a smart electromagnetic environment will be enabled by non-invasive, cognitive, low-power and low-complexity intelligent metasurfaces (MTSs) to be integrated within, or on top of, objects, machines and/or structural building elements. MTSs are artiﬁcial surfaces constituted by electrically small elements that collectively exhibit equivalent homogeneous boundary conditions to any interacting electromagnetic ﬁelds. In the years 2000-2010, MTSs for radio frequency (RF) applications were essentially uniform in space and realized by periodically printing elements on a dielectric slab. This was the ﬁrst generation of MTSs. In a second generation (2010-2020) MTSs are designed to create spatially variable boundary conditions to further tailor the interaction with the incident ﬁeld. Today, we are facing a transition to the third generation of MTSs, where MTSs change boundary conditions in both space and time in intelligent way, opening new perspectives for the next generation of communications. This section of the roadmap focuses on envisioning surface wave-based sensing MTSs as a complement to the re-routing MTSs, in the context of an intelligent connectivity environment.

\paragraph{Introduction}
The awareness capability of a smart radio-environment consists on the ability to capture the instantaneous electromagnetic (EM) status by using a set of independent EM sensorial surfaces connected through a controlled electronic network, whose aim is to perceive the localization of the network players (e.g., access-points and users). The instantaneous amplitude and phase of the EM field is transduced into electrical quantities at individual ports connected by an electronic intelligence behind. 

\begin{figure}[ht]\centering
	\includegraphics[width=0.6\linewidth]{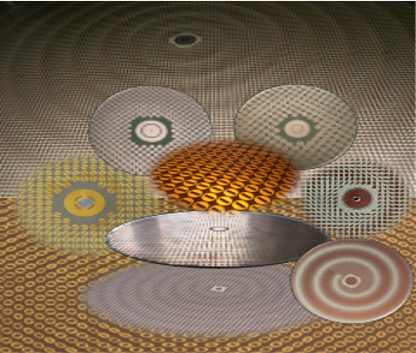}
	\caption{MTS antennas based on SWs. The MTS is formed by a texture of patches printed on a grounded dielectric slab.}
	\label{fig1M}
\end{figure}

\begin{figure}[thb]\centering
	\includegraphics[width=0.6\linewidth]{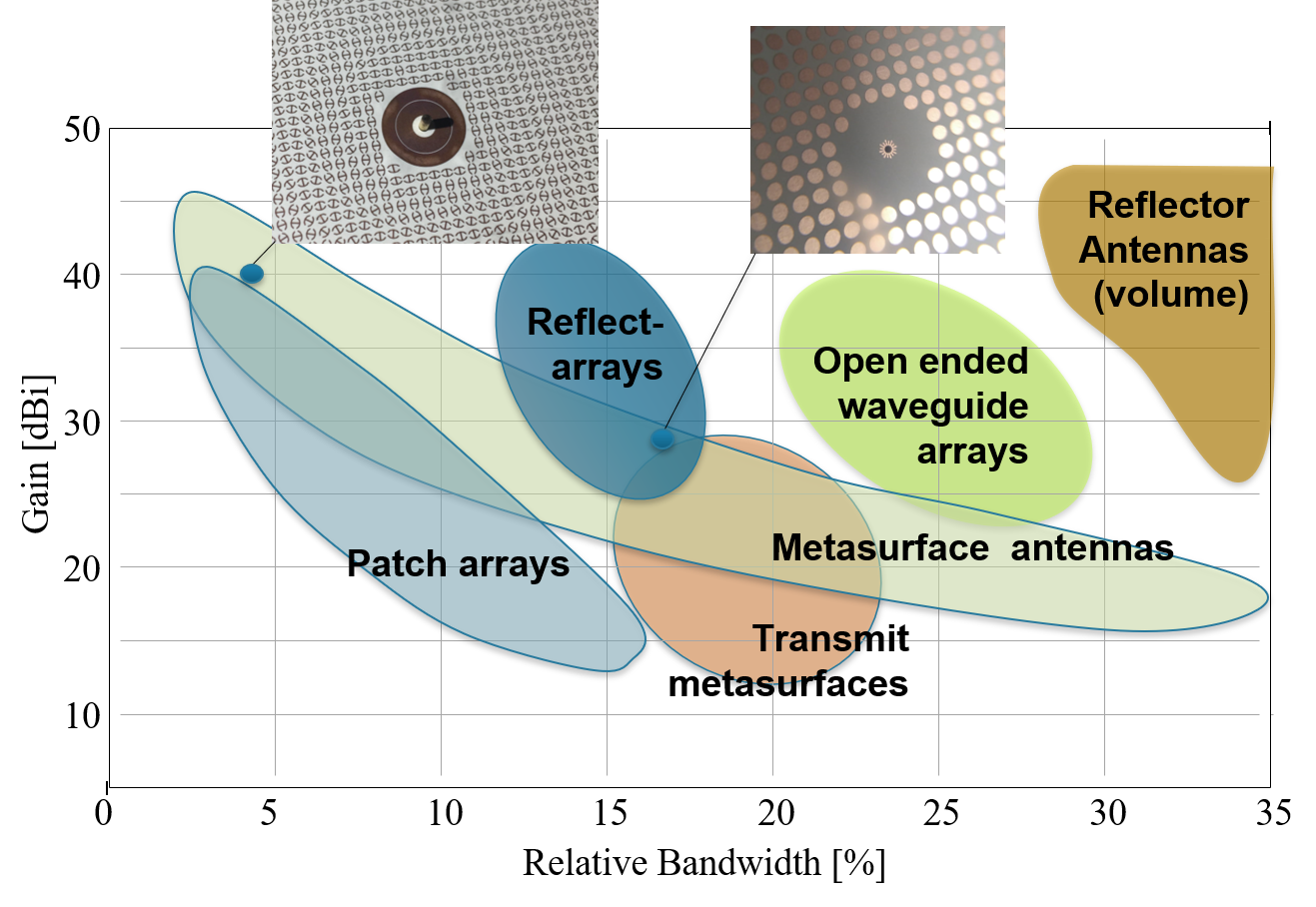}
	\caption{MTS antenna area in a GB diagram in comparison with other types of PCB antennas. The insets show the layout of antennas realized with high-gain and small bandwidth (40dB, 5\%) and large product bandwidth-gain (30 dB, 17\%)}
	\label{fig2M}
\end{figure}

The simplest EM receiving smart device is a sensing MTS, properly emulating impedance boundary conditions (IBCs) that convert EM space wave from the environment into a surface wave (SW), whose energy is collected at one or more points on the surface. In literature, these devices are called SW-based modulated MTSs \cite{Holographic}.  
At microwaves, a SW-based MTS is composed of a grounded dielectric slab with many small metallic printed elements, often referred to as “pixels”, arranged in a regular lattice (Figure \ref{fig1M}). The SW is collected by an elementary sensor embedded in the MTS itself, that can be as simple as an elementary monopole. MTS receiving antennas have therefore a simple, low profile structure.  This mechanism is the reciprocal of the one for which a Leaky Wave (LW) originates from a monopole-excited cylindrical SW which interacts with periodic or quasi-periodic  IBCs \cite{space}. 
By changing the geometrical characteristics of the inclusions, the surface can be adapted to receive external fields from different arrival directions, or from multiple simultaneous directions \cite{Multibeam}, and arbitrary polarization \cite{Amagoia}. This feature has fostered the research on dynamically adaptive MTS, which are able to steer or reconfigure the beams to reach the maximum reception at the sensor point/points \cite{Smith}. The main concept is to dynamically change the IBCs offered by the MTS by acting on the inclusions composing the MTS through active electronic devices or tunable materials \cite{Sievenpiper}. 
SW-based MTSs are cost effective, easily manufactured with standard PCB techniques, light, low profile and conformable to different kinds of physical surfaces. These are intriguing features for smart environment connectivity.  Moreover, they can be applied on a wide range of fractional-bandwidth (B) and gain (G). Figure \ref{fig2M} shows the region occupied by MTS antennas in a G-B diagram, in comparison with other types of antennas. 

The most recent achievement is the large bandwidth performance,  i.e. pattern stability in frequency  \cite{Wideband}. Pattern bandwidth is limited by the dispersion-induced mismatch between the SW wavelength and the periodicity of the IBCs occurring when the operational frequency changes. By using a non-uniform modulation period one can enlarge the bandwidth, at the price of a lower, but still significant, antenna efficiency. This habilitates the designer to play with high gain and reduced bandwidth or viceversa, thus, spanning the large area of the diagram in Fig. \ref{fig2M}.  
Overall, the bandwidth performance of these antennas are quite robust with respect to alternative printed dipole technology (i.e. single substrate patch arrays, reflectarrays, and transmitarray) with single feed-point. Furthermore, reflectarrays or transmit-MTSs are not really flat because of the presence of an  external feeder. 

\paragraph{Emerging challenges}
The main challenge for intelligent MTSs able to localize the network players (e.g., access-points and users) is represented by the complexity of the electronics. Two different architectures can be conceived to minimize it. The first one is based on a dynamic configuration with active control switches. The second one is based on a static configuration with overlapped holographies and multiple receiving points. The two architectures are described next. 
\\

{\it A. Uniform MTS with controllable switches and single pin} 

Any electrically small constituent element of the MTS is connected to a switch that allows to change the particle status with two bits (zero-one). The distribution of switches in status “on” is changed continuously maintaining a density of half-wavelength. The distribution is able to re-design a holography of “on” switches capable of sensing one direction of arrival at a time, so as the SW coupled energy will be maximized at the single collector point (\cite{Sievenpiper, Smith}). The inclusions of the MTS are loaded with active devices or they include phase changing materials, such as liquid crystals or vanadium dioxide. The electric features of the inclusions become voltage controlled, and, hence, the IBCs offered by the MTS can be properly adjusted by an external control. However, this solution  needs to be improved in terms of efficiency. In fact, active devices or phase changing materials yield an increase of the antenna losses, with a consequent reduction of gain and increase of power demand; losses become more important when working at higher frequencies (e.g. Ka-band): frequency scalability is indeed another future challenge. Also, phase changing materials suffer of temperature instability. Alternative strategies make use of optical pumping of silicon or gallium arsenide substrates to alter the electrical properties of the inclusions. Losses become less critical when dealing with a mechanical reconfiguration. The use of micromechanical systems or piezoelectric devices have been proposed, but they may suffer of low reliability and an intelligent MTS based on such devices might be too sensitive to the external vibrations if installed on moving vehicles. 
\\

{\it B. Multi-pin overlapped holographies}

It has been demonstrated \cite{Multibeam} that a MTS can be designed in such a way to simultanously couple waves coming from different directions maximizing power at individual collector points. This is done by a proper optimization of overlapping contributions that maximize the power received from different directions. This approach exhibits low-complexity, since no switches are needed, and it offers the possibility of detecting simultaneously more directions of arrival (DoA), distinguishing them by ports. However, the main drawback with respect to the solution A is the smaller effective area of collection for each DoA, which translates in a reduction of precision in the DoA detection. 

\paragraph{Future developments to satisfy these challenges}
Advances in phase changing materials would bring a significant benefit to MTS antennas in terms of efficiency, switching speed and temperature stability. MTSs with integrated active devices, instead of discrete active devices, would also bring advantages in terms of reliability, losses and performance. Realizing integrated devices on a wide area, in turn, will require an improvement of the accuracy of the realization processes. 

\paragraph{Conclusion}
SW based MTS antennas constitute an innovative concept for sensing the EM smart environment in non-invasive, cognitive, and low-complexity manner. Space-time tailoring of MTSs is obtained with active devices or phase changing materials.


\section*{\fontsize{20}{25}\selectfont Waveform-Selective Metasurfaces and Their Potential Applications in Wireless Communications} 
\subsection*{Hiroki Wakatsuchi\textsuperscript{1,2}* and Sendy Phang\textsuperscript{3}} 
\subsubsection*{\textsuperscript{1}\textit{Department of Electrical and Mechanical Engineering, Graduate School of Engineering, Nagoya Institute of Technology, Nagoya, Aichi, 466-8555, Japan}} 
\subsubsection*{\textsuperscript{2}\textit{Precursory Research for Embryonic Science and Technology (PRESTO), Japan Science and Technology Agency (JST), Saitama 332-0012, Japan}} 
\subsubsection*{\textsuperscript{2}\textit{George Green Institute of Electromagnetics Research, Faculty of Engineering, University of Nottingham, Nottingham NG7 2RD, UK}} 
\subsubsection*{*\textbf{Corresponding author}: wakatsuchi.hiroki@nitech.ac.jp} 
\vspace{5mm}

\paragraph{Introduction}
In this Roadmap, we introduce recently developed waveform-selective metasurfaces \cite{Reference1,Reference2} and explain how they potentially fit in emerging issues/technologies in wireless communications. Composed of subwavelength conducting elements connected by lumped circuits, waveform-selective metasurfaces are capable of sensing a particular incident wave among others even at the same frequency depending on their waveforms or, more specifically, on their pulse widths. Such a waveform selectivity is made possible by coupling electromagnetic resonant mechanisms of metasurfaces with transient phenomena well known in classic direct-current (DC) circuits. Specifically, diodes are used to rectify electric charges induced on conductors and to generate an infinite set of frequency components including zero frequency. The energy converted to zero frequency is then controlled by other lumped circuit components such as capacitors and inductors over a time period much longer than a cycle of an incident wave (see more detail in the literature  \cite{Reference1,Reference3,Reference4}). Waveform-selective metasurfaces were so far reported to preferentially sense various types of waveforms \cite{Reference3,Reference5} and vary their absorptances \cite{Reference1,Reference3,Reference6}, scattering parameters \cite{Reference4,Reference7}, and polarization changes \cite{Reference4} at a normal/oblique angle \cite{Reference8,Reference9} with/without tuning systems \cite{Reference4}. Besides conventional frequency selectivity, such a waveform selectivity provides an additional degree of freedom to address existing electromagnetic issues as discussed below.

\begin{figure}[ht]\centering
	\includegraphics[width=0.95\linewidth]{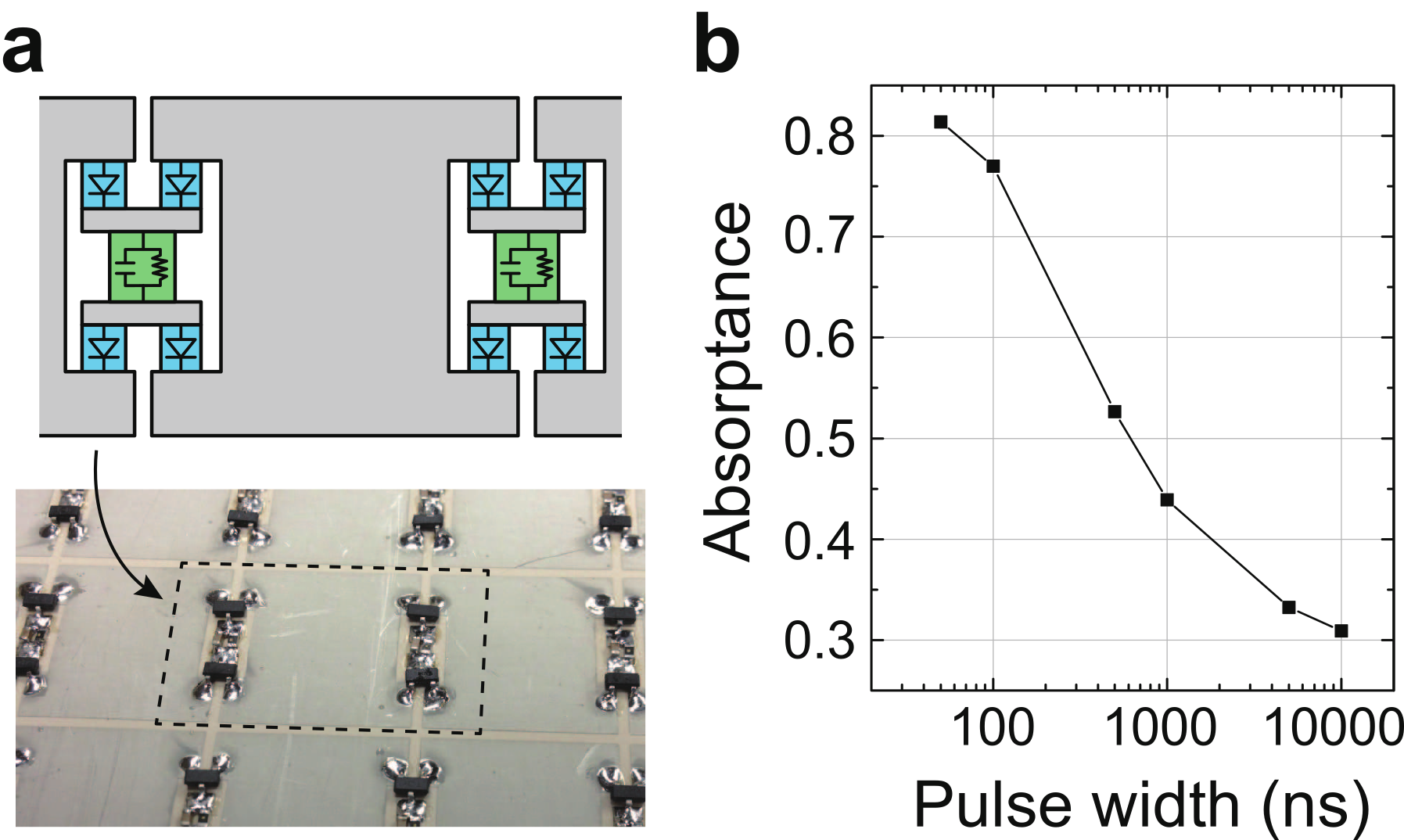}
	\caption{(a) Waveform-selective metasurface and (b) its absorbing profile at the “same” frequency of 4.2 GHz \cite{Reference1}. }
	\label{fig1}
\end{figure}

\paragraph{Applications proposed in the literature}
A metallic enclosure composed of shielding conductor walls can be used to protect sensitive electronic devices from strong electromagnetic fields. Such an enclosure, however, is known to generate stronger internal fields than external ones due to its cavity resonance \cite{Reference10}. This electromagnetic interference issue can be readily mitigated by including an absorbing material inside the enclosure \cite{Reference11}. However, this solution also weakens communication signals generated from the inside of the cavity. A conducting cavity composed of a waveform-selective transmitting metasurface simultaneously achieves a strong shielding effect for a continuous wave and an enhanced transmission effect that rather exploits the cavity resonance to transmit a short pulse for a longer distance even at the same frequency \cite{Reference4}. In addition, waveform-selective metasurfaces can be used to directly cover an antenna. This scheme can be used to eliminate a coupling between individual antenna elements and design advanced intelligent array systems  \cite{Reference12}. Another interesting application reported so far is seen in signal processing where a large reflective screen consists of different types of pixelated waveform-selective metasurfaces \cite{Reference13}. By mathematically solving the relationship between time-varying multiple inputs and outputs, an incident signal can be restored as a temporal ghost imaging. We also emphasise that waveform-selective metasurfaces can control not only fundamental electromagnetic properties but also communication characteristics such as bit-error-rates \cite{Reference3,Reference14}. In the following part of this Roadmap, we particularly discuss how these metasurfaces are aligned with emerging issues/technologies in wireless communications.

\paragraph{Potential applications in wireless communications}
The number of wireless communication devices keeps increasing in modern society. For instance, internet of things (IoT) devices are newly installed at a pace of more than two billion a year over the world \cite{Reference15}. While these devices enrich the quality of our life, a new technique is necessary for accommodating such a vast number of devices within a single network and ensuring concurrent connection. In this context, various multiaccess and multiplexing techniques were proposed using frequency, space, time, etc as their modulation variables \cite{Reference16,Reference17}. Waveform selectivity can potentially add a new degree of freedom that is conceptually orthogonal to any of these variables and thus contributes to enhancing network capacities and connection performances. Another issue seen here is how advanced IoT systems composed of an extremely large number of sensors are maintained if the sensors are highly dispersed in space. In this case, ideally each sensor is expected to be free of a battery for less maintenance and collect its driving energy from surrounding electromagnetic fields \cite{Reference18}. Waveform-selective metasurfaces store rectified energy in internal circuit components such as capacitors, which can empower IoT sensors. 

Also, the Fifth generation (5G) of mobile communications provide high-speed communication services by using a quasi-millimeter wave band, which at the same time narrows a coverage area, compared to that of a lower frequency band around a few GHz. A promising solution to this issue is introducing wavefront-shaping panels that reflect incident waves to non-line-of-sight devices \cite{Reference19}. Although such a reflection profile usually remains unchanged, a metasurface combining a wavefront-shaping functionality with waveform selectivity enables dynamic steering, thereby further increasing a coverage area. Moreover, waveform selectivity can be more usefully exploited together with cloaks in the generation of the post 5G, where the main frequency domain is supposed to be in the THz regime. Especially, for indoor scenarios, surface waves may be more often used to effectively send signals to distant receivers than ever before as free-space wave propagation has a limited travelling distance range. However, a practical design of wireless communication environment involves inevitable scattering objects or columns. These scatterers can be bypassed by using a cloak. Moreover, if the cloaking system is composed of waveform-selective metasurfaces and contains an antenna, waveform-selective communications and cloaking are simultaneously realised.

\paragraph{Conclusion}
In conclusion, we have briefly covered recent studies on waveform-selective metasurfaces that vary their responses depending on incident waveforms or pulse widths. We explained their fundamental characteristics and applications with a strong focus on wireless communications. Nevertheless, the concept of waveform-selective metasurfaces is simple and versatile to be applied to other existing electromagnetic issues as well.

\phantomsection
\section*{Acknowledgment}
H.W. acknowledges support from the Japan Science and Technology Agency (JST) under Precursory Research for Embryonic Science and Technology (PRESTO) No. JPMJPR193A and the Japanese Ministry of Internal Affairs and Communications (MIC) under Strategic Information and Communications R\&D Promotion Program (SCOPE) No. 192106007.


\end{document}